\documentclass[pra, aps, twocolumn, groupedaddress, showpacs, superscriptaddress]{revtex4-2}
\usepackage{amssymb,amsmath,amsthm,color,graphicx,times,graphicx}
\usepackage{hyperref}
\usepackage{subfigure}
\usepackage{times}
\usepackage{bbold}
\usepackage{color}
\usepackage{orcidlink}

\usepackage{subfigure}

\usepackage{tikz}
\usetikzlibrary{shapes}
\usetikzlibrary{shapes}

\usepackage{hyperref}
\hypersetup{
	colorlinks=true,
	linkcolor=blue,
	filecolor=blue,      
	urlcolor=blue,
	citecolor=violet,
	pdfauthor={Noureddine Abouelkhir, Abdallah Slaoui, Hanane El Hadfi and Rachid Ahl Laamara},
	pdftitle={Estimating phase parameters of a three-level system interacting with two classical monochromatic fields in simultaneous and individual metrological strategies},
	pdfcreator={Abdallah Slaoui},
}

\providecommand{\openone}{\leavevmode\hbox{\small1\kern-4.3pt\normalsize1}}

\theoremstyle{plain}

\theoremstyle{definition}

\begin{document}

\title{Estimating phase parameters of a three-level system interacting with two classical monochromatic fields in simultaneous and individual metrological strategies}
\author{Nour-Eddine Abouelkhir \orcidlink{0000-0002-6164-0525}}\affiliation{LPHE-Modeling and Simulation, Faculty of Sciences, Mohammed V University in Rabat, Rabat, Morocco.}
\author{Abdallah Slaoui \orcidlink{0000-0002-5284-3240}}\email{Corresponding author: abdallah.slaoui@um5s.net.ma
}\affiliation{LPHE-Modeling and Simulation, Faculty of Sciences, Mohammed V University in Rabat, Rabat, Morocco.}\affiliation{Centre of Physics and Mathematics, CPM, Faculty of Sciences, Mohammed V University in Rabat, Rabat, Morocco.}
\author{Hanane El Hadfi}\affiliation{LPHE-Modeling and Simulation, Faculty of Sciences, Mohammed V University in Rabat, Rabat, Morocco.}
\author{Rachid Ahl Laamara}
\affiliation{LPHE-Modeling and Simulation, Faculty of Sciences, Mohammed V University in Rabat, Rabat, Morocco.}\affiliation{Centre of Physics and Mathematics, CPM, Faculty of Sciences, Mohammed V University in Rabat, Rabat, Morocco.}
\begin{abstract}
Recently, the Hilbert-Schmidt speed, as a special class of quantum statistical speed, has been reported to improve the interferometric phase in single-parameter quantum estimation. Here, we test this concept in the multiparameter scenario where two laser phases are estimated in a theoretical model consisting of a three-level atom interacting with two classical monochromatic fields. When the atom is initially prepared in the lower bare state taking into account the detuning parameters, we extract an exact analytical solution of the atomic density matrix in the case of two-photon resonant transition. Further, we compare the performance of laser phase parameters estimation in individual and simultaneous metrological strategies, and we explore the role of quantum coherence in improving the efficiency of unknown multi-phase shift estimation protocols. The obtained results show that the Hilbert-Schmidt speed detects the lower bound on the statistical estimation error as well as the optimal estimation regions, where its maximal corresponds to the maximal quantum Fisher information, the performance of simultaneous multiparameter estimation with individual estimation inevitably depends on the detuning parameters of the three-level atom, and not only the quantum entanglement, but also the quantum coherence is a crucial resource to improve the accuracy of a metrological protocol.

\end{abstract}
\date{\today}

\maketitle

\section{Introduction}
During the last two decades, the use of quantum systems to process information has given rise to a new field of research called quantum information, which has developed rapidly \cite{Gabor1950,Watrous2018}. At the same time, advances in the experimental field now allow the controlled manipulation of many quantum systems \cite{Holevo2011,Helstrom1969}. These two aspects have contributed to the development of new technologies, which rely on the use of quantum resources, such as quantum entanglement \cite{Schrodinger1935,Einstein1935,Makouri2023} or quantum superposition \cite{Rouse1995,Nakamura1999}, to make the processes implemented more efficient. On the other hand, the process of measurement, which aims to assign a value to a physical quantity when producing an estimate, is one of the most modern technological applications of the physical sciences. Every experimental estimate is accompanied by an uncertainty that has an impact on the result of the measurement, whether it is fundamental or technological. In fact, physical laws impose fundamental restrictions on uncertainty such as those induced by the Heisenberg relations. Additionally, the technical error is mainly represented by an involuntary error resulting from uncontrollable defects in the measurement procedure. Since quantum physics is the most predictive theory and allows us to describe phenomena that have no equivalent in classical physics, it is appropriate to study, within the framework of this theory, the measurement process as well as the limitations of the achievable precision \cite{SlaouiDrissi2022,Bakraoui2022}.\par

The quantum technology revolution, which aspires to create brand-new technologies that take advantage of quantum phenomena, has prompted a recent development of quantum estimation theory. It has been used to construct high-precision measurements at the quantum frontier by providing theoretical tools for a variety of estimation objectives. These applications include clock synchronization \cite{Joza2010}, optimal estimation of phases \cite{Ballester2004,Abouelkhir2023,Aspachs2009}, Unruh-Hawking effect estimation \cite{Aspachs2010}, space-time parameters \cite{Nation2009}, reservoirs temperature \cite{Correa2015}, the sensitivity of gravitational wave interferometry \cite{Schnabel2010} and standard frequency estimation \cite{Boss2017}. Indeed, quantum metrology provides the tools to obtain estimates of quantities of interest for a physical system but not accessible by direct measurement \cite{Braunstein1994,Paris2009}. Among these quantities, there are quantities that do not correspond to quantum observables such as the quantum phase, the purity of a quantum state and quantum correlations. In particular, the considerations made in this framework are quantified by calculating the quantum Fisher information (QFI) (Fisher information (FI) in the classical case) which allows to define the precision of the estimation of a physical parameter present in the Hamiltonian of the system. According to the Cramer-Rao bound (CRB) ${\rm Var}(\hat{\theta})\geq {\cal F}_\theta^{-1}$, where ${\rm Var}(\hat{\theta})$ is the variance of the estimator $\hat{\theta}$ and ${\cal F}_\theta$ denotes the FI quantity \cite{Cramer1999}, the FI provides a lower bound on the uncertainty of the parameter estimate. The QFI provides an upper bound on the FI for any measurement and, therefore, the best possible estimation precision is given by quantum mechanics. Besides, a small variance provides the highest precision of the estimated parameter. Hence, the ultimate goal of any quantum estimation process is to achieve the smallest variance value. In this frame, it turned out that the optimal way to estimate a phase shift in the presence of phase-diffusion and to obtain the ultimate quantum limits of accuracy is to take into account schemes with noise as well as to consider the environmental effects in the optical phase measurement \cite{Genoni2011}. Single-parameter unitary gates estimation, optimal probes and the role of entanglement to improve the overall stability of the estimation scheme (i.e. the robustness of the optimal settings with respect to fluctuations of the probe and the measurement parameters), for qubit systems using Bayesian inference are explored in ref \cite{Teklu2009}. Moreover, contrary to separable states, the advantage of using entangled qubits as quantum probes to characterize the noise induced by an external complex environment and to improve the estimation of the correlation time has been pointed out in \cite{Rossi2015}. It is also interesting to note that, in realistic scenarios, the improvement in quantum precision in practical applications of phase estimation tasks is bounded by the phase diffusion \cite{Brivio2010}. Besides, in Ref.\cite{Teklu2010} the same improvement has been proved for a single qubit undergoing an unknown phase shift imposed by the unitary dynamics of quantum state.\par

As previously mentioned, the main key to quantum estimation theory is the QCRB which always reaches saturation in the case where a single parameter is estimated. On the other hand, it is difficult to saturate this bound when many parameters are estimated simultaneously, in which the variance is replaced by covariance matrix ${\rm Cov}(\hat{\theta})$ and QFI by the QFI matrix $F_{\theta}$ (i.e., ${\rm Cov}(\hat{\theta})\geq F_{\theta}^{-1}$). This results from the incompatibility between the measurements of the different estimated parameters. For this, multiparameter quantum estimation has attracted a lot of interest and has become an important task in a variety of diverse settings. Further, when multiple variables are estimated simultaneously, in a way that can outperform individual estimation strategies, simultaneous estimation can provide better precision than their individual estimation. The myriad reasons for studying multi-parameter quantum estimation schemes are deeply intertwined and many theoretical and experimental studies have been conducted \cite{Szczykulska2016}, among them decoherence parameters estimation \cite{Crowley2014}, linear and nonlinear phase shift estimation using two-mode entangled coherent states \cite{Cheng2014}, and  estimating multiple-parameter unitary operators \cite{Kolenderski2008,Bagan2001,Vaneph2013}. Moreover, the investigation of noise and dissipative effects on several open quantum systems, in which the bosonic or fermionic character of their constituents requires a reformulation of the standard concepts of information theory, becomes of paramount importance. The research work in this area is not only of theoretical but also of practical interest, as experimental developments in the field of quantum system control have paved the way for metrological applications in which the presence of coherence and entanglement allows experimental accuracies in the estimation of some parameters otherwise inaccessible \cite{Benatti2010,Braun2018}.\par
On the other hand, quantum optics is nowadays one of the liveliest fields of physics; lately, it has experienced a very rapid development thanks to the progress of experimental techniques that allow the creation and detection of the photon, as well as the investigation of an atom inside the cavities. Partly motivated by these advances, several theoretical and experimental works have explored many quantum phenomena in these optical systems, such as the enhancement of entanglement performance and the optical bistability of a two- or even three-mode continuous-valued system, for example in a cavity optomechanical system that studies the interplay between the light field and the mechanical motion by the radiation pressure. Indeed, continued interests have been focused on entanglement generation via the single-atom system and its investigation for two moving mirrors coupled to a two-mode laser inside a doubly resonant cavity was done in \cite{Teklu2018}, and then in Ref.\cite{Wang2022} where one-way steering and stable entanglement in a single atom at four-levels interacting simultaneously with two cavity modes are generated.\par
Emerging applications, such as ion traps, superconducting qubits, and quantum dots, among others, are driving the exploration of the collective properties of these quantum systems \cite{Allen1997,Li2019,Steane1997,Michler2017}. Different strategies to describe the interaction of these systems with the external field were proposed and one of the most important models is the Jaynes-Cummings model (JCM) \cite{Jaynes1963}. It describes a system composed of a quantized mode interacting with a two-level system, and was originally proposed to describe the interaction in a strongly idealized way between a single atom and a single mode of the electromagnetic field, both isolated from the influence of any perturbing environment. The JCM is of great interest for atomic physics \cite{Sukumar1981,Shaukat2020,SlaouiShaukat2018,Schlicher1989}, quantum optics \cite{Larson2022} and quantum information processing \cite{Blais2020,Devoret2013}, both theoretically and experimentally. Furthermore, this model has been extended to additional directions, such as adding more levels where three-level atomic systems under different configuration; lamda ($\Lambda$), Cascade ($\Xi$) and Vee ($V$), have been examined \cite{Li1987}. There is also evidence for the effects of atomic motion \cite{Schlicher1989}, the Kerr-like medium \cite{Agarwal1989}, Stark-shift \cite{Nasreen1993,Slaoui2020} and multi-phonon atomic transitions \cite{Sukumar1981}. Inspired by these works, we investigate here simultaneous and individual metrological strategies to improve the estimation of phase parameters for a three-level atom interacting with laser fields in the case of a two-photon resonant transition. The atomic density matrix under the rotating wave approximation is obtained. Besides, we employ the concept of Hilbert-Schmidt speed, as a quantum statistical speed quantifier, to detect the accuracy of the estimated phase parameters. The roles of quantum coherence and geometric phase in multi-parameter quantum estimation are also addressed.\par

This paper is arranged as follows: In  Section \ref{Sec2}, we present the Hamiltonian describing the interaction between a three-level atom and two classical monochromatic fields. By using the Schrödinger equation, the atomic density matrix when the atom is prepared in the lowest energy bare state is obtained. The construction of the QFI matrix, the simultaneous and individual estimation methods, and the bound on the joint estimation of the phase parameters are all provided in Section \ref{Sec3}. Importantly, the QFI matrix is not diagonal and it and the optimal bound depend on the phase parameters. In Section \ref{Sec4}, we devoted ourselves to the roles of the Hilbert-Schmidt speed as well as the quantum coherence in the precision of the multiparametric quantum estimation. We derive here their analytical expressions and compare their behaviors with those of the QFI. Finally, conclusions and some features and comments are given.

\section{Physical model and its atomic density matrix}\label{Sec2}

\begin{figure}[hbtp]
	\centering
	\includegraphics[scale=0.18]{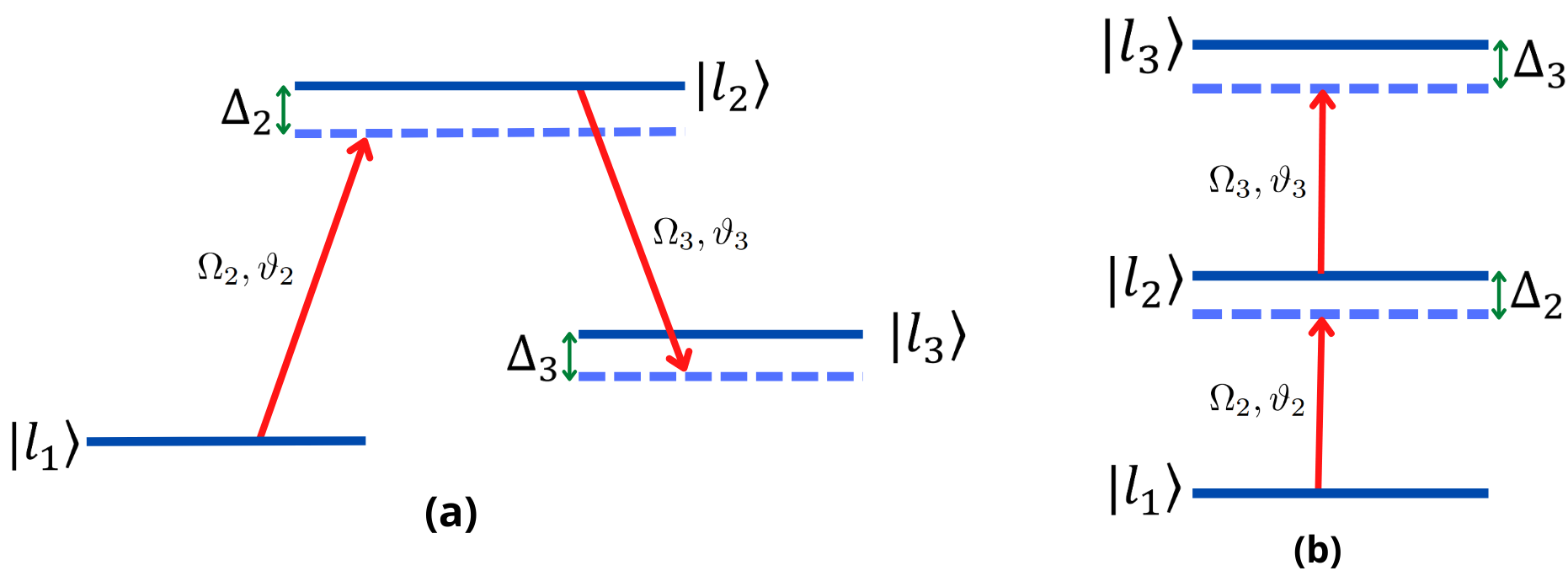} 
	\caption{The energy levels diagrams of a three-level atom subject to two classical monochromatic fields, with detuning parameters $\Delta_{2}$ and $\Delta_{3}$ for ({\bf a}) $\Lambda$-configuration and ({\bf b}) $\Xi$-configuration.}\label{three}
\end{figure}

The theoretical model to be considered is a idealized three-level atom interacting with two classical monochromatic fields. As shown schematically in Fig.(\ref{three}), two levels $|l_{2}\rangle$ and $|l_{3}\rangle$ are coupled to a single level $|l_{1}\rangle$ on two dipole-allowed transitions driven by fields at frequencies $\Omega_{2}$ and $\Omega_{3}$, respectively. Assume that there are two possible configurations of the bare states ($\Xi$ and $\Lambda$-type) with two lower levels $|l_{2}\rangle$ and $|l_{3}\rangle$ are coupled to a single upper level $|l_{1}\rangle$. Besides, the $E_{2}$ field causing the transition $|l_{1}\rangle\to|l_{2}\rangle$ has a phase $\phi_{2}$ and is detuned from resonance by a frequency $\Delta_{2}$. The $E_{3}$ field causing the transition $|l_{2}\rangle\to|l_{3}\rangle$ has a phase $\phi_{3}$ and is detuned by frequency $\Delta_{3}$  (see Fig.(\ref{three})).

The Hamiltonian of the aforementioned model is given by $\hat{H}=\hat{H}_{0}+\hat{V}$, where
\begin{equation}
\hat{H}_{0}=\hbar\omega_{2}|l_{2}\rangle\langle l_{2}|+\hbar\omega_{3}|l_{3}\rangle\langle l_{3}|,
\end{equation}
is the unperturbed Hamiltonian and the time-dependent perturbation is given by
\begin{equation}\label{V}
\begin{aligned}
\hat{V}&=-\hat{\mu}_{2}E_{2}-\hat{\mu}_{3}E_{3}\\
&=-\mu_{2}E_{2}(|l_{1}\rangle\langle l_{2}|+|l_{2}\rangle\langle l_{1}|)-\mu_{3}E_{3}\left(|l_{2}\rangle\langle l_{3}|+|l_{3}\rangle\langle l_{2}|\right),
\end{aligned}
\end{equation}
with $\hat{\mu}_{2}$ (resp. $\hat{\mu}_{3}$) is the atomic dipole operator associated with the two states $|l_{1}\rangle$ and $|l_{2}\rangle$ (resp. $|l_{2}\rangle$ and $|l_{3}\rangle$). For a monochromatic field, the electric fields can be written as 
\begin{equation}
E_{2}=E_{02}\cos\left(\Omega_{2} t+\phi_{2}\right) \quad \mbox{and} \quad E_{3}=E_{03}\cos\left(\Omega_{3} t+\phi_{3}\right).\label{eq3}
\end{equation}
Reporting this last equation (\ref{eq3}) into the equation (\ref{V}), this leads us to the definition of a key parameter for atom-light interactions. The Rabi frequencies associated with the coupling of the field modes are given by
\begin{equation}
\vartheta_{2}=\mu_{2}E_{02}/\hbar, \quad \mbox{and} \quad \vartheta_{3}=\mu_{3}E_{03}/\hbar.\label{eq4}
\end{equation}
Afterwards, using the rotating-wave approximation (RWA), the associated Hamiltonian of the atom-field system can be written in terms of projection operators as
\begin{equation}
\begin{aligned}
\hat{H}&=\hbar\omega_{2}|l_{2}\rangle\langle l_{2}|+\hbar\omega_{3}|l_{3}\rangle\langle l_{3}|\\
&-\frac{\hbar}{2}\vartheta_{2}\left(e^{i(\Omega_{2} t+\phi_{2})}|l_{1}\rangle\langle l_{2}|+e^{-i(\Omega_{2} t+\phi_{2})}|l_{2}\rangle\langle l_{1}|\right)\\
&-\frac{\hbar}{2}\vartheta_{3}\left(e^{i(\Omega_{2} t+\phi_{2})}|l_{2}\rangle\langle l_{3}|+e^{-i(\Omega_{3} t+\phi_{3})}|l_{3}\rangle\langle l_{2}|\right).
\end{aligned}
\end{equation}
Since there are two laser frequencies in our system, the usual unitary transformation to the rotating frame at $\vartheta$ does not apply. However, it is possible to perform a unitary transformation that makes $\hat{H}$ time independent, i.e. $|\psi(t)\rangle_{I}={\cal U}^{\dagger}|\psi(t)\rangle$ (rotating frame) where ${\cal U}=\sum_{j=1}^{3}\exp\left[-i\gamma_{j}|l_{j}\rangle\langle l_{j}|\right]$. In the interaction picture, the Hamiltonian takes the following form
\begin{equation}
\hat{H}_{I}={\cal U}^{\dagger}\hat{H}{\cal U}+i\hbar \frac{\partial {\cal U}^{\dagger}}{\partial t}{\cal U},
\end{equation}
Hence, the Hamiltonian can rewritten as follows
\begin{widetext}
\begin{equation}\label{gg}
\hat{H_{I}}=\hbar\left[\begin{array}{cccc}
-\gamma_{1} & -\frac{1}{2}\vartheta_{2}e^{i(\Omega_{2}+\gamma_{1}-\gamma_{2})t+\phi_{2})} & 0 \\
-\frac{1}{2}\vartheta_{2}e^{-i(\Omega_{2}+\gamma_{1}-\gamma_{2})t+\phi_{2})} & \omega_{2}-\gamma_{2} & -\frac{1}{2}\vartheta_{3}e^{i(\Omega_{3}-\gamma_{2}+\gamma_{3})t+\phi_{3})} \\
0 & -\frac{1}{2}\vartheta_{3}e^{-i(\Omega_{3}-\gamma_{2}+\gamma_{3})t+\phi_{3})} & \omega_{3}-\gamma_{3} 
\end{array}\right].
\end{equation}
\end{widetext}
Next, our goal is to determine the appropriate values of $\gamma_{j}$ for making our Hamiltonian independent of time. For this purpose, we can choose the following values for $\gamma_{j}$'s:
\begin{equation}
\Omega_{2}+\gamma_{1}-\gamma_{2}=0,\quad \Omega_{3}-\gamma_{2}+\gamma_{3}=0, \quad \mbox{and} \quad -\gamma_{1}=0,
\end{equation}
from this, we can conclude that
\begin{equation}
\gamma_{2}=\Omega_{2},\quad \gamma_{3}=\Omega_{2}-\Omega_{3}, \quad \mbox{and} \quad \gamma_{1}=0.
\end{equation}
The time-independent Hamiltonian of the system in the rotating-wave approximation can be written in terme of projection operators as
\begin{equation} \label{H}
\begin{aligned}
\hat{H_{I}}=& \hbar \Delta_{2}|l_{2}\rangle\langle l_{2}|+\hbar(\Delta_{2}+\Delta_{3})|l_{3}\rangle\langle l_{3}| \\
& -\frac{\hbar}{2}\left(\vartheta_{2}e^{-i\phi_{2}}|l_{2}\rangle\langle l_{1}|+\vartheta_{3}e^{i\phi_{3}}|l_{2}\rangle\langle l_{3}|+h.c\right),
\end{aligned}
\end{equation}
where the energy of the bare state $|l_{1}\rangle$ is zero. In our consideration, we assume that the interaction occurs wherein the energy of the bare state $|l_{1}\rangle$ comes to zero. The dressed states are the eigenstates of this Hamiltonian (\ref{H}) which are denoted by $|i\rangle$, $|j\rangle$, and $|k\rangle$. 
These dressed states are superpositions of the bare states $|l_{A}\rangle$, $|l_{2}\rangle$ and $|l_{3}\rangle$, as
\begin{equation}\label{super}
|a\rangle=\alpha_{a}|l_{1}\rangle+\beta_{a}|l_{2}\rangle+\gamma_{a}|l_{3}\rangle, \quad a=i,j,\mbox{ or }k,
\end{equation}
where $\alpha_{a}$, $\beta_{a}$ and $\gamma_{a}$ are complex constants. We require these states to be eigenstates satisfying the equations
\begin{equation}\label{Evp}
\hat{H}|a\rangle=\hbar\omega_{a}|a\rangle.
\end{equation}
Putting the equations (\ref{H}) and (\ref{super}) into (\ref{Evp}), the eigenvalue equation can be written as
\begin{align}
&\left[-\frac{\vartheta_{2}}{2} e^{i\phi_{2}} \beta_{a}|l_{1}\rangle+\left(-\frac{\vartheta_{2}}{2} e^{-i\phi_{2}} \alpha_{a}+\Delta_{2} \beta_{a}-\frac{\vartheta_{3}}{2} e^{i\phi_{3}} \gamma_{a}\right)|l_{2}\rangle\right.\notag \\&
\left.+\left(-\frac{\vartheta_{3}}{2} e^{-i\phi_{3}} \beta_{a}+\left(\Delta_{2}+\Delta_{3}\right) \gamma_{a}\right)|l_{3}\rangle\right]=\notag \\&\omega_{a}\left(\alpha_{a}|l_{1}\rangle+\beta_{a}|l_{2}\rangle+\gamma_{a}|l_{3}\rangle\right).
\end{align}
By comparing the coefficients of the bare states, we obtain three simultaneous equations for $\alpha_{a}$, $\beta_{a}$ and $\gamma_{a}$ which can be written in matrix form as follows
\begin{equation}\label{M}
\left(\begin{array}{ccc}
-\omega_{a} & -\frac{\vartheta_{2}}{2} e^{i\phi_{2}} & 0 \\
-\frac{\vartheta_{2}}{2} e^{-i\phi_{2}} & \Delta_{2}-\omega_{a} & -\frac{\vartheta_{3}}{2} e^{i\phi_{3}} \\
0 & -\frac{\vartheta_{3}}{2} e^{-i\phi_{3}} & \Delta_{2}+\Delta_{3}-\omega_{a}
\end{array}\right)\left(\begin{array}{c}
\alpha_{a} \\
\beta_{a} \\
\gamma_{a}
\end{array}\right)=0.
\end{equation}
Indeed, a non-trivial solution for $\alpha_{a}$, $\beta_{a}$ and $\gamma_{a}$ requires a zero determinant of the above matrix (\ref{M}). This leads to
\begin{equation}\label{cubic Eq} \small
\begin{aligned}
 \omega_{a}^{3}-\omega_{a}^{2}(2\Delta_{2}&+\Delta_{3})+\omega_{a}\left(\Delta_{2}(\Delta_{2}+\Delta_{3})-\frac{\vartheta_{2}^{2}}{4}-\frac{\vartheta_{3}^{2}}{4}\right) \\
 & +\frac{\vartheta_{2}^{2}}{4}(\Delta_{2}+\Delta_{3})=0.
\end{aligned}
\end{equation}
Here we consider the situation in which the bare states $|l_{1}\rangle$ and $|l_{3}\rangle$ have a two-photon resonant transition, in the sense that $\Delta_{2}+\Delta_{3}=0$. If we write $\Delta_{2}=-\Delta_{3}=\Delta$, the cubic equation (\ref{cubic Eq}) reads
\begin{equation}\label{E2}
\omega_{a}^{3}-\Delta\omega_{a}^{2}-\frac{1}{4}(\vartheta_{2}^{2}+\vartheta_{3}^{2})\omega_{a}=0,
\end{equation}
and thus, the eigenfrequencies are provided by
\begin{equation}\label{Vp}
\begin{aligned}
&\omega_{i}=\frac{1}{2}\left(\Delta-\sqrt{\Delta^{2}+\vartheta_{2}^{2}+\vartheta_{3}^{2}}\right), \\
&\omega_{j}=0, \\
&\omega_{k}=\frac{1}{2}\left(\Delta+\sqrt{\Delta^{2}+\vartheta_{2}^{2}+\vartheta_{3}^{2}}\right).
\end{aligned}
\end{equation}
Due to the fact that $\Delta_{2}=-\Delta_{3}$ and the selected interaction picture, in which the bare state $|l_{1}\rangle$ has zero energy, the two-photon resonance case has zero eigenfrequency. The normalized eigenvectors of the matrix (\ref{M}) with these eigenfrequencies can be expressed as follows
\begin{equation} \small
|i\rangle=\frac{1}{\sqrt{4\omega_{i}^{2}+\vartheta_{2}^{2}+\vartheta_{3}^{2}}}\left(-\vartheta_{2}e^{i\phi_{2}}|l_{1}\rangle+2\omega_{i}|l_{2}\rangle-\vartheta_{3} e^{-i\phi_{3}}|l_{3}\rangle\right),
\end{equation}
and
\begin{equation} \small
|k\rangle=\frac{1}{\sqrt{4\omega_{k}^{2}+\vartheta_{2}^{2}+\vartheta_{3}^{2}}}\left(-\vartheta_{2}e^{i\phi_{2}}|l_{1}\rangle+2\omega_{k}|l_{2}\rangle-\vartheta_{3} e^{-i\phi_{3}}|l_{3}\rangle\right).
\end{equation}
Thus, it is easier to find the eigenvector corresponding to the zero eigenfrequency from the Hamiltonian matrix (\ref{M}) with $\omega_{a}=\omega_{j}=0$. Starting from the resultant matrix, we observe that the normalized eigenvector is
\begin{equation}
|j\rangle=\frac{1}{\sqrt{\vartheta_{2}^{2}+\vartheta_{3}^{2}}}\left(-\vartheta_{3}e^{i\phi_{3}}|l_{1}\rangle+\vartheta_{2} e^{-i\phi_{2}}|l_{3}\rangle\right).\label{eq11}
\end{equation}
In fact, the dressed state above (\ref{eq11}) is the eigenvector corresponding to the zero eigenfrequency and it is a superposition of only the bare states $|l_{1}\rangle$ and $|l_{3}\rangle$ and does not include any contribution from the bare state $|l_{2}\rangle$. A sum over the dressed states determines the system's evolved state so that
\begin{equation}\label{State}
|\psi(t)>=C_{i}e^{-i\omega_{i}t}|i>+C_{j}|j>+C_{k}e^{-i\omega_{k}t}|k>,
\end{equation}
where $C_{a}$ are complex amplitudes calculated from the initial state of the atom. We assume the atom to be initially in its lower bare state, so that $|\psi(0)>=|l_{1}>$. Under such condition we have
\begin{equation}
|l_{1}\rangle=C_{i}|i\rangle+C_{j}|j\rangle+C_{k}|k\rangle,
\end{equation}
with
\begin{equation}\label{Ca}
\begin{aligned}
&C_{i}=\langle i|l_{1}\rangle=\frac{-1}{\sqrt{4 \omega_{a}^{2}+\vartheta_{2}^{2}+\vartheta_{3}^{2}}}\vartheta_{2}e^{-i\phi_{2}}, \\
&C_{j}=\langle j|l_{1}\rangle=\frac{-1}{\sqrt{\vartheta_{2}^{2}+\vartheta_{3}^{2}}}\vartheta_{3}e^{-i\phi_{3}}, \\
&C_{k}=\langle k|l_{1}\rangle=\frac{-1}{\sqrt{4 \omega_{c}^{2}+\vartheta_{2}^{2}+\vartheta_{3}^{2}}}\vartheta_{2}e^{-i\phi_{2}}.
\end{aligned}
\end{equation}
In the interesting limiting case, the detuned $\Delta$ is much larger than either $\vartheta_{2}$ or $\vartheta_{3}$. Therefore, the eigenfrequencies are derived by expanding (\ref{Vp}) to be
\begin{equation}
\omega_{i}\simeq -\frac{(\vartheta_{2}^{2}+\vartheta_{3}^{2})}{4\Delta}, \quad \omega_{j}=0, \quad \omega_{k}\simeq\Delta,
\end{equation}
with $\omega_{i}$ holds an lower order of magnitude than both $\vartheta_{2}$ and $\vartheta_{3}$, and $\omega_{j}$ holds an higher order. Moreover, according to (\ref{Ca}), we get
\begin{equation}
C_{i}\simeq \frac{-\vartheta_{2}e^{-i\phi_{2}}}{\sqrt{\vartheta_{2}^{2}+\vartheta_{3}^{2}}}, \hspace{1cm} C_{j}\simeq \frac{-\vartheta_{3}e^{-i\phi_{3}}}{\sqrt{\vartheta_{2}^{2}+\vartheta_{3}^{2}}},
\end{equation}
however $C_{k}$, which is of order $1/\Delta$, may be disregarded. Using the aforementioned findings with this order of approximation, the state (\ref{State}) is a superposition of just two dressed states provided by
\begin{equation}\label{eq17}
|\psi(t)\rangle\simeq \frac{-1}{\sqrt{\vartheta_{2}^{2}+\vartheta_{3}^{2}}}\left[\vartheta_{2}e^{-i\phi_{2}}e^{-i\omega_{i}t}|i\rangle+\vartheta_{3}e^{-i\phi_{3}}|j\rangle\right].
\end{equation}
After the derivation of the atomic density matrix operator, $\rho=\left|\psi(t) \right\rangle \left\langle\psi(t)\right|$, we can calculate the corresponding QFI matrix to estimate the phase parameters $\phi_{2}$ and $\phi_{3}$ as well as we will investigate the simultaneous and individual estimation strategies for the considered system.
\section{Quantum Multiparameter Estimation Theory}\label{Sec3}
Effectively, estimation theory refers to the branch of science that examines the accuracy with which a given set of physical parameters can be estimated. In multi-parameter estimation scenarios, multiple variables are evaluated simultaneously and the theoretical limits on their sensitivity have become possible through the FI matrix, which quantifies the amount of information contained in a data set about unknown parameters, and its related Cramér-Rao bound \cite{Paris2009,Cramer1999}. To estimate certain parameters $\boldsymbol{\theta}=\left\lbrace\theta_{1},\theta_{2},...,\theta_{n}\right\rbrace $ that is an open subset of ${\cal R}^{n}$ and encrypted in the dynamics $\varphi_{\theta}$, with $\boldsymbol{\hat{\theta}}=\left\lbrace\hat{\theta}_{1},\hat{\theta}_{2},...,\hat{\theta}_{n}\right\rbrace$ being an estimator of $\boldsymbol{\theta}$ and $\hat{\theta}_{\nu}$  the estimator of $\theta_{\nu}$, first prepare a probe state $\rho$ and allow it to evolve under the evolution $\rho\xrightarrow{\varphi_{\theta}}\rho_{\theta}$ , and then, by conducting POVM measurements $\left\lbrace \Pi_{x}\right\rbrace$ on the output state $\rho_{\theta}$, the result of measurement $x$ with the probability $p(x/\theta)={\rm Tr}\left(\Pi_{x}\rho_{\theta}\right)$ is obtained. According to the Cramér-Rao theorem which asserts that for any measure $\Pi_{x}$ and an unbiased estimator $\hat{\theta}_{\nu}$, the precision of the parameter estimate is bounded by
\begin{equation}\label{CRB}
N{\rm Cov}\left(\boldsymbol{\hat{\theta}}\right)\geq {\cal I}^{-1}\left(\boldsymbol{\theta}\right),
\end{equation}
where $N$ is the number of independent measurements, ${\rm Cov}(\hat{\theta})$ stands for the covariance matrix of all unbiased estimators $\hat{\theta}$ whose elements are ${\rm Cov}(\hat{\theta})_{\mu\nu}=E\left[\theta_{\mu}\theta_{\nu}\right]-E\left[\theta_{\mu}\right]E\left[\theta_{\nu}\right]$, $E$ being a mathematical expectation, and ${\cal I}\left(\boldsymbol{\theta}\right)$ is the classical Fisher information matrix with the $\mu\nu$th entry, in terms of the distribution function $p(x/\theta)$, given by 
\begin{equation}\label{CFI}
{\cal I}\left(\boldsymbol{\theta}\right)_{\mu\nu}=\int p(x/\theta)\left(\frac{\partial \log p(x/\theta)}{\partial\theta_{\mu}}\frac{\partial \log p(x/\theta)}{\partial\theta_{\nu}}\right)dx.
\end{equation}
Here, the attainability conditions and the optimal estimators are those where the inequality CCRB is saturated. Moreover, the maximum estimator is found to be optimal in the limit of a large number of measurements $N\longmapsto\infty$. An upper quantum limit on the CFI matrix is provided for a quantum system by the fluctuation of the unknown parameters in the system state $\rho_{\theta}$, which allows to obtain the quantum Cramér-Rao inequality
\begin{equation}\label{QCRB}
N{\rm Cov}\left(\boldsymbol{\hat{\theta}}\right)\geq {\cal I}^{-1}\left(\boldsymbol{\theta}\right)\geq\mathcal{F}^{-1}\left(\boldsymbol{\theta}\right),
\end{equation}
where the $\mu\nu$th entry of quantum Fisher information matrix $\mathcal{F}^{-1}\left(\boldsymbol{\theta}\right)$ is given by
\begin{equation} \label{QFIM}
\mathcal{F}_{\mu\nu}\left(\theta\right)=\frac{1}{2}{\rm Tr}\left[\rho_{\theta}(\hat{L}_{\theta_{\mu}}\hat{L}_{\theta_{\nu}}+\hat{L}_{\theta_{\nu}}\hat{L}_{\theta_{\mu}})\right],
\end{equation}
which depends only on the final state $\rho_{\theta}$ since the measurement $\Pi_{x}$ was automatically optimized on all POVMs, where the symmetric logarithmic derivative (SLD) $\hat{L}_{\theta_{\mu}}$ (with $\mu=1,..,n$) satisfies the following differential equation
\begin{equation}\label{SLD}
\frac{\partial\rho_{\theta}}{\partial\theta_{\mu}}=\frac{1}{2}\{\hat{L}_{\theta_{\mu}}\rho_{\theta}+\rho_{\theta}\hat{L}_{\theta_{\mu}}\}.
\end{equation}

The above equation needs to be solved to derive the SLDs $\hat{L}_{\theta_{\mu}}$ corresponding to the various estimated parameters. For a pure state $\rho_{\theta}^{2}\equiv\rho_{\theta}$, which interests us here, solving the equation (\ref{SLD}) gives us the expression of SLD as $\hat{L}_{\theta_{\mu}}=2\partial_{\theta_{\mu}}\rho_{\theta}$ and then the QFI matrix elements become
\begin{equation}\label{QFIM pur}
\mathcal{F}_{\mu\nu}=4Re\left[\langle\partial_{\theta_{\mu}}\psi|\partial_{\theta_{\nu}}\psi\rangle-\langle\partial_{\theta_{\mu}}\psi|\psi\rangle\langle\psi|\langle\partial_{\theta_{\nu}}\psi\rangle\right].
\end{equation}
According to the QCR bound, one may be able to estimate the parameter $\theta_{\mu}$ by making an appropriate measurement more precisely with a smaller statistical error, when the QFI matrix element $\mathcal{F}_{\mu\mu}$ is larger. Typically, the quantum Cramer-Rao inequality (\ref{QCRB}) become scalar inequalities and it is always saturable by optimizing over all valid quantum measurements in single-parameter estimation scenarios, i.e ${\rm Var}(\theta)\geq \mathcal{F}^{-1}_{\theta\theta}$. By using projectors on the eigenvectors of the SLD operators $L_{\theta}$, this saturation yields the optimal quantum measurement operators. To the contrary, the matrix Cramer-Rao inequality in multiparameter estimation protocols, ${\rm Cov}(\boldsymbol{\hat{\theta}})\geq \mathcal{F}^{-1}(\boldsymbol{\theta})$, is usually not saturable. Indeed, the measurements made by the optimal operators for various parameters may not be compatible. Hence, quantum bounds on precision are not generally achieved. In order to saturate the bound (\ref{QCRB}), we should be aware of the influence of this incompatibility on our estimation problem. In fact, one can find a common eigenbasis for all SLDs in the case where the $L_{\theta}$ operators commute. This implies that we can perform a simultaneous measurement saturating the QCR inequality \cite{Ragy2016,Bakmou2019,Napoli2019}. In the situation where the SLDs are not commuted, the condition ${\rm Tr}\{\hat{\rho}[\hat{L}_{\theta_{\mu}},\hat{L}_{\theta_{\nu}}]\}=0$ for $\forall(\theta_{\mu},\theta_{\nu})\in\boldsymbol{\theta}$ is sufficient for saturating the QCR bound. This commutation condition is reduced for pure states to
\begin{equation}\label{eq24}
{Im}\langle\partial_{\theta_{\mu}}\psi|\partial_{\theta_{\nu}}\psi\rangle=0.
\end{equation}
Generally speaking, the QCR bound can be attainable if a matrix ${\cal U}_{\theta_{\nu}\theta_{\mu}}=-\frac{i}{2}{\rm Tr}(\rho[\hat{L}_{\theta_{\mu}},\hat{L}_{\theta_{\nu}}])$ called Uhlmann curvature matrix with elements in equation (\ref{eq24}) vanishes, thus the influence of the incompatibility condition on multiparameter estimation problems is quantified by the quantity ${\cal R}_{\boldsymbol{\theta}}:=\|2i\mathcal{F}^{-1}\left(\boldsymbol{\theta}\right){\cal U}_{\boldsymbol{\theta}}\|_{\infty}$ referred to as quantumness, where $\|{\bf B}\|_{\infty}$ stands for the largest eigenvalue of the matrix ${\bf B}$ \cite{Candeloro2021,Albarelli2020}. Importantly, the maximum incompatibility between the measures associated with the estimated parameters is equivalent to the saturation of the upper bound, and in this case the quantumness ${\cal R}_{\boldsymbol{\theta}}=1$. This quantity has been used to evaluate quantitatively the critical phenomena of the geometric phase for many-body systems interacting with critical chains in non-equilibrium phase transitions \cite{Carollo2018}. Further, recent theoretical contributions based on the  Uhlmann curvature matrix have suggested superior accuracy and stability performance when using quantum probes \cite{CandeloroRazavian2021} and when using coherently driven nonlinear Kerr resonators \cite{Asjad2023} over their classical counterparts.\par

In our considered model described by equation (\ref{eq17}), where we want to estimate the phase parameters of the external fields $\phi_{2}$ and $\phi_{3}$, it is simple to verify that the condition (\ref{eq24}) is fulfilled by taking $\theta_{\mu}\equiv\phi_{2}$ and $\theta_{\nu}\equiv\phi_{3}$. Instead, it is straightforward to check that $[\hat{L}_{\phi_{2}},\hat{L}_{\phi_{3}}]=0$, which does not imply any indeterminacy due to the quantum compatibility arising from the quantum nature of the parameters estimation problem. Thus, the optimal values of the two estimated parameters are mutually compatible and then have a common optimal basis. The eigenvalues of such SLD operators provide the optimal states and therefore their projections are the optimal measurements. Therefore, the QCR bound is always saturable and the optimal accuracy is obtained when the QFI of the estimated parameter is maximized.
\subsection{\bf Individual Metrological Strategy}
We examine here the best strategies for extracting the information about the phase parameters of two classical monochromatic fields $\phi_{2,3}$ coupled to a three level atom. In the single-parameter protocol, where the optimal scenario is the one that produces the minimum estimation variance, the non-zero QFI matrix elements are found only along the diagonal which are given by
\begin{equation} \label{QFI single pur}
\mathcal{F}_{\theta\theta}=4\left[\langle\partial_{\theta}\psi|\partial_{\theta}\psi\rangle-|\langle\partial_{\theta}\psi|\psi\rangle|^{2}\right].
\end{equation}
In this strategy in which we estimated the phase parameters individually, these parameters are statistically independent and the accurate identification of a single phase does not affect the precision of the other, so that $\mathcal{F}_{\phi_{2}\phi_{3}}$. Therefore, the QCR inequalities become
\begin{equation}
{\rm Var}\left(\phi_{2}\right)_{\rm Ind}\geq\mathcal{F}^{-1}_{\phi_{2}},\hspace{0.5cm}{\rm and}\hspace{0.5cm}{\rm Var}\left(\phi_{3}\right)_{\rm Ind}\geq\mathcal{F}^{-1}_{\phi_{3}}.\label{eq26}
\end{equation}
Equation (\ref{eq26}) show the existence of an optimal probe state containing a maximum amount of QFI with exact values of the estimated phase parameters, for which the best accuracy is provided. After some straightforward algebra, the analytical expressions of the QFI for the laser phase $\phi_{2}$ is
\begin{equation}\label{R QFI single}
\mathcal{F}_{\phi_{2}}=4\left(4\vartheta_{2}^{2}\kappa+\vartheta_{2}^{2}\vartheta_{3}^{2}\xi\right)\left[1-\left(4\vartheta_{2}^{2}\kappa+\vartheta_{2}^{2}\vartheta_{3}^{2}\xi\right)\right],
\end{equation}
and the QFI for the laser phase $\phi_{3}$ is written as
\begin{equation}
\mathcal{F}_{\phi_{3}}=4\vartheta_{2}^{2}\vartheta_{3}^{2}\xi\left(1-\vartheta_{2}^{2}\vartheta_{3}^{2}\xi\right),
\end{equation}
with the quantities $\kappa$ and $\xi$ are given by
\begin{align}
\kappa=&\frac{\omega_{i}^{2}}{(4\omega_{i}^{2}+\vartheta_{2}^{2}+\vartheta_{3}^{2})^{2}}+\frac{\omega_{k}^{2}}{(4\omega_{k}^{2}+\vartheta_{2}^{2}+\vartheta_{3}^{2})^{2}} \notag \\
& +\frac{2\omega_{i}\omega_{k}\cos((\omega_{i}-\omega_{k})t)}{(4\omega_{i}^{2}+\vartheta_{2}^{2}+\vartheta_{3}^{2})(4\omega_{k}^{2}+\vartheta_{2}^{2}+\vartheta_{3}^{2})},\label{eq29}
\end{align}
and
\begin{align}
\xi&=\frac{1}{(4\omega_{i}^{2}+\vartheta_{2}^{2}+\vartheta_{3}^{2})^{2}}+\frac{1}{(4\omega_{k}^{2}+\vartheta_{2}^{2}+\vartheta_{3}^{2})^{2}} \notag \\&+\frac{2\cos((\omega_{i}-\omega_{k})t)}{(4\omega_{i}^{2}+\vartheta_{2}^{2}+\vartheta_{3}^{2})(4\omega_{k}^{2}+\vartheta_{2}^{2}+\vartheta_{3}^{2})}+\frac{1}{(\vartheta_{2}^{2}+\vartheta_{3}^{2})^{2}}\notag \\ & \quad -\frac{2}{\vartheta_{2}^{2}+\vartheta_{3}^{2}}\left(\frac{\cos(\omega_{i}t)}{4\omega_{i}^{2}+\vartheta_{2}^{2}+\vartheta_{3}^{2}}+\frac{\cos(\omega_{k}t)}{4\omega_{k}^{2}+\vartheta_{2}^{2}+\vartheta_{3}^{2}}\right).\label{eq30}
\end{align}
\subsection{\bf Simultaneous Metrological Strategy}
We can now address the problem of generalizing this treatment to the simultaneous multi-parameter strategy, where we aim to measure these two physical quantities $\phi_{2,3}$ simultaneously and seeking to get as close as possible to the highest possible accuracy. Actually, a simultaneous estimation strategy requires less quantum resources than the individual estimation technique. If the estimated parameters are compatible, none of the parameters loses accuracy but the required resources (e.g. coherence, squeezing, entanglement or energy in the input state preparation) are thus reduced, resulting in the greatest possible improvement. In order to describe uncertainty of estimators $\boldsymbol{\hat{\theta}}\equiv\left\lbrace\phi_{2},\phi_{3}\right\rbrace$ in the simultaneous metrological schemes, the covariance matrix ${\rm Cov}(\boldsymbol{\hat{\theta}})$ can be renewed as
\begin{equation}
{\rm Cov}\left(\boldsymbol{\hat{\theta}}\right) =\left[\begin{array}{ccc}
{\rm Var} \left( \phi_{2}\right)&{\rm Cov} \left( \phi_{2},\phi_{3}\right)\\
{\rm Cov} \left( \phi_{3},\phi_{2}\right) & {\rm Var} \left( \phi_{3}\right) 
\end{array}\right],\label{eq31}
\end{equation}
and the inverse of QFI matrix can be written straightly as
\begin{equation}
\mathcal{F}^{-1}\left(\boldsymbol{\theta}\right) =\frac{1}{{\rm det}\left(\mathcal{F}\left(\boldsymbol{\theta}\right) \right)}\left[\begin{array}{ccc}
\mathcal{F}_{\phi_{3}}&-\mathcal{F}_{\phi_{2}\phi_{3}} \\
-\mathcal{F}_{\phi_{3}\phi_{2}} & \mathcal{F}_{\phi_{2}}
\end{array}\right],\label{eq32}
\end{equation}
with ${\rm det}\left(\mathcal{F}\left(\boldsymbol{\theta}\right) \right)=\mathcal{F}_{\phi_{2}}\mathcal{F}_{\phi_{3}}-\mathcal{F}_{\phi_{2}\phi_{3}}^{2}$. Replacing equations (\ref{eq31}) and (\ref{eq32}) in the inequality (\ref{QCRB}) which are both positive semidefinite, and using Sylvester's criterion \cite{Prussing1986}, we obtain
\begin{equation}
{\rm Var}\left( \phi_{2}\right)_{\rm Sim}\geq\frac{\mathcal{F}_{\phi_{3}}}{{\rm det}\left(\mathcal{F}\left(\boldsymbol{\theta}\right) \right)}, \hspace{0.5cm}{\rm Var}\left(\phi_{3}\right)_{\rm Sim}\geq\frac{\mathcal{F}_{\phi_{2}}}{{\rm det}\left(\mathcal{F}\left(\boldsymbol{\theta}\right) \right)},
\end{equation}
and the important trade-off between ${\rm Var}\left( \phi_{\mu,\nu}\right)$, $\mathcal{F}_{\phi_{\mu,\nu}}$ and ${\rm Cov}\left(\phi_{\mu},\phi_{\nu}\right)$ is captured by 
\begin{align}
&\left({\rm Var}\left( \phi_{2}\right)_{\rm Sim}- \frac{\mathcal{F}_{\phi_{3}}}{{\rm det}\left(\mathcal{F}\left(\boldsymbol{\theta}\right) \right)}\right)\left({\rm Var}\left( \phi_{3}\right)_{\rm Sim}- \frac{\mathcal{F}_{\phi_{2}}}{{\rm det}\left(\mathcal{F}\left(\boldsymbol{\theta}\right) \right)}\right)\notag\\&\geq\left({\rm Cov}\left(\phi_{2},\phi_{3}\right)- \frac{\mathcal{F}_{\phi_{2}\phi_{3}}}{{\rm det}\left(\mathcal{F}\left(\boldsymbol{\theta}\right) \right)}\right).
\end{align}
By exploiting the equation (\ref{QFIM pur}), one can easily show that
\begin{equation}
\mathcal{F}_{\phi_{2}\phi_{3}}=\mathcal{F}_{\phi_{3}\phi_{2}}=4\vartheta_{2}^{2}\vartheta_{3}^{2}\xi\left[1-\left(4\vartheta_{2}^{2}\kappa+\vartheta_{2}^{2}\vartheta_{3}^{2}\xi\right)\right],
\end{equation}
in term of the quantities $\kappa$ (eq.\ref{eq29}) and $\xi$ (eq.\ref{eq30}). To compare the performance of the estimates of the phase parameters in the individual and simultaneous schemes, we use here a technique whereby we introduce the ratio between the total variances in these two strategies as follows
\begin{equation}\label{ratio}
\gamma=\frac{\Delta_{\rm Ind}}{\Delta_{\rm Sim}},
\end{equation}
with the total variance in the individual case is
\begin{equation}
\Delta_{\rm Ind}={\rm Var}\left(\phi_{2}\right)_{\min}^{\rm Ind}+{\rm Var}\left(\phi_{3}\right)_{\min}^{\rm Ind},
\end{equation}
and in the simultaneous case become 
\begin{equation}
\Delta_{\rm Sim}=\frac{1}{2}\left[{\rm Var}\left(\phi_{2}\right)_{\min}^{\rm Sim}+{\rm Var}\left(\phi_{3}\right)_{\min}^{\rm Sim}\right].
\end{equation}
Here it should be mentioned that the minimal values of the variances for estimating parameter ($\phi_{2},\phi_{3}$) individually are
\begin{equation}
{\rm Var}\left(\phi_{2}\right)_{\min}^{\rm Ind}=\mathcal{F}_{\phi_{2}}^{-1},\hspace{0.5cm} {\rm Var}\left(\phi_{3}\right)_{\min}^{\rm Ind}=\mathcal{F}_{\phi_{3}}^{-1},
\end{equation}
and for the simultaneous estimation are
\begin{equation}\label{minimal Var}
\rm {Var}\left(\phi_{2}\right)_{\min}^{\rm Sim}=\frac{\mathcal{F}_{\phi_{3}}}{{\rm det}\left(\mathcal{F}\left(\boldsymbol{\theta}\right) \right)},\hspace{0.5cm}{\rm Var}\left(\phi_{3}\right)_{\min}^{\rm Sim}=\frac{\mathcal{F}_{\phi_{2}}}{{\rm det}\left(\mathcal{F}\left(\boldsymbol{\theta}\right) \right)}.
\end{equation}
Reporting these last equations into equation (\ref{ratio}), the ratio $\gamma$ can then be written rather elegantly as
\begin{equation}
\gamma=\frac{2{\rm det}\left(\mathcal{F}\left(\boldsymbol{\theta}\right) \right)}{\mathcal{F}_{\phi_{2}}\mathcal{F}_{\phi_{3}}}.
\end{equation}
In the total variance formula $\Delta_{\rm Sim}$ we inserted a factor $1/2$, since we estimated two parameters simultaneously. This factor is necessary to take into account the reduction in resources which showed that the simultaneous strategy requires $2$ less resources than the individual scheme in the multiparameter estimation procedures. As a result, $\gamma\leq2$ generally, and when the ratio $\gamma>1$, the error limit of the simultaneous parameter estimation scheme is smaller and offers an advantage over that of the individual case.\par

\begin{widetext}
	
		\begin{figure}[hbtp]
		{{\begin{minipage}[b]{.5\linewidth}
					\centering
					\includegraphics[scale=0.4]{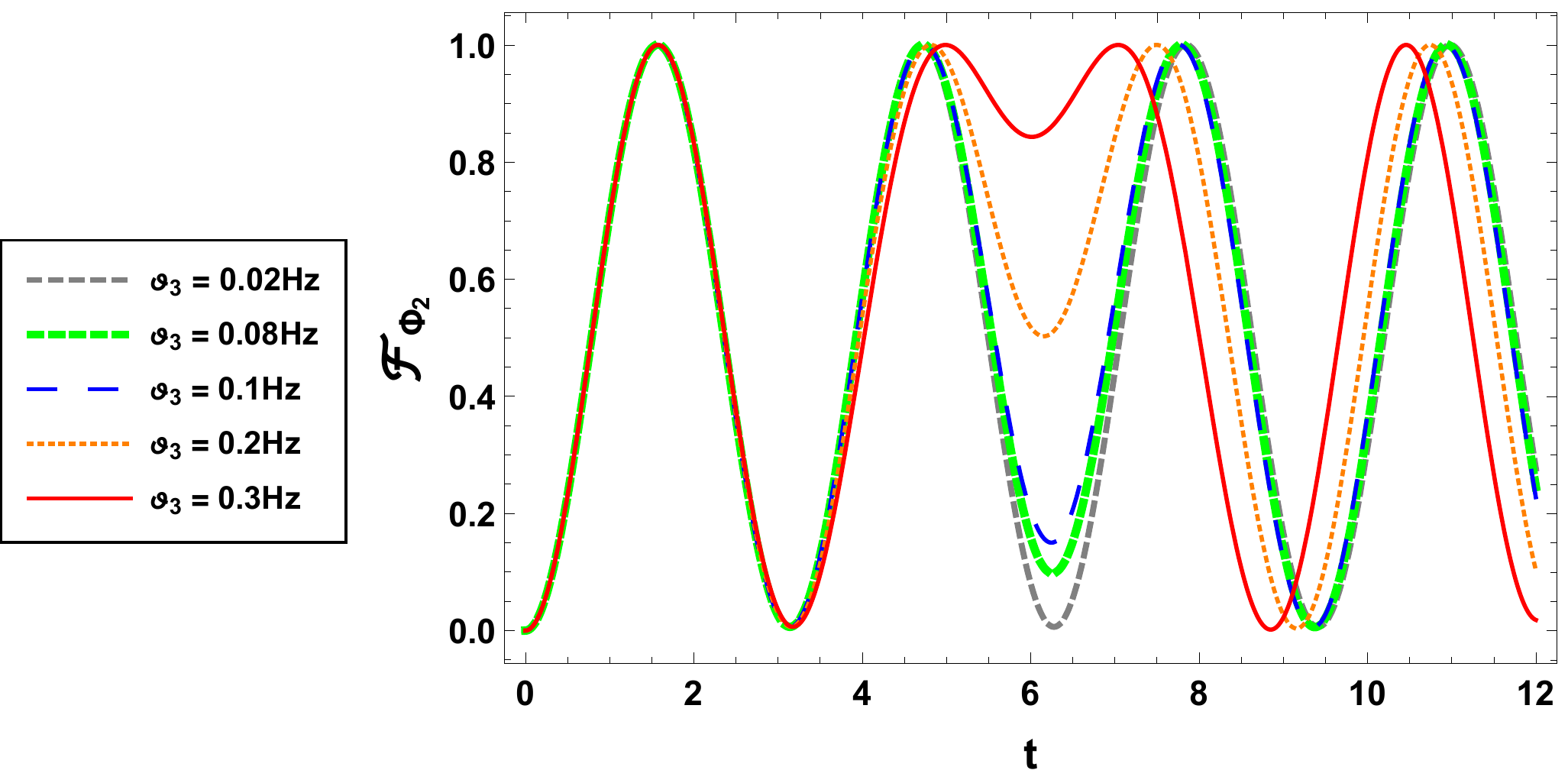} \vfill $\left(a\right)$
				\end{minipage}\hfill
				\begin{minipage}[b]{.5\linewidth}
					\centering
					\includegraphics[scale=0.4]{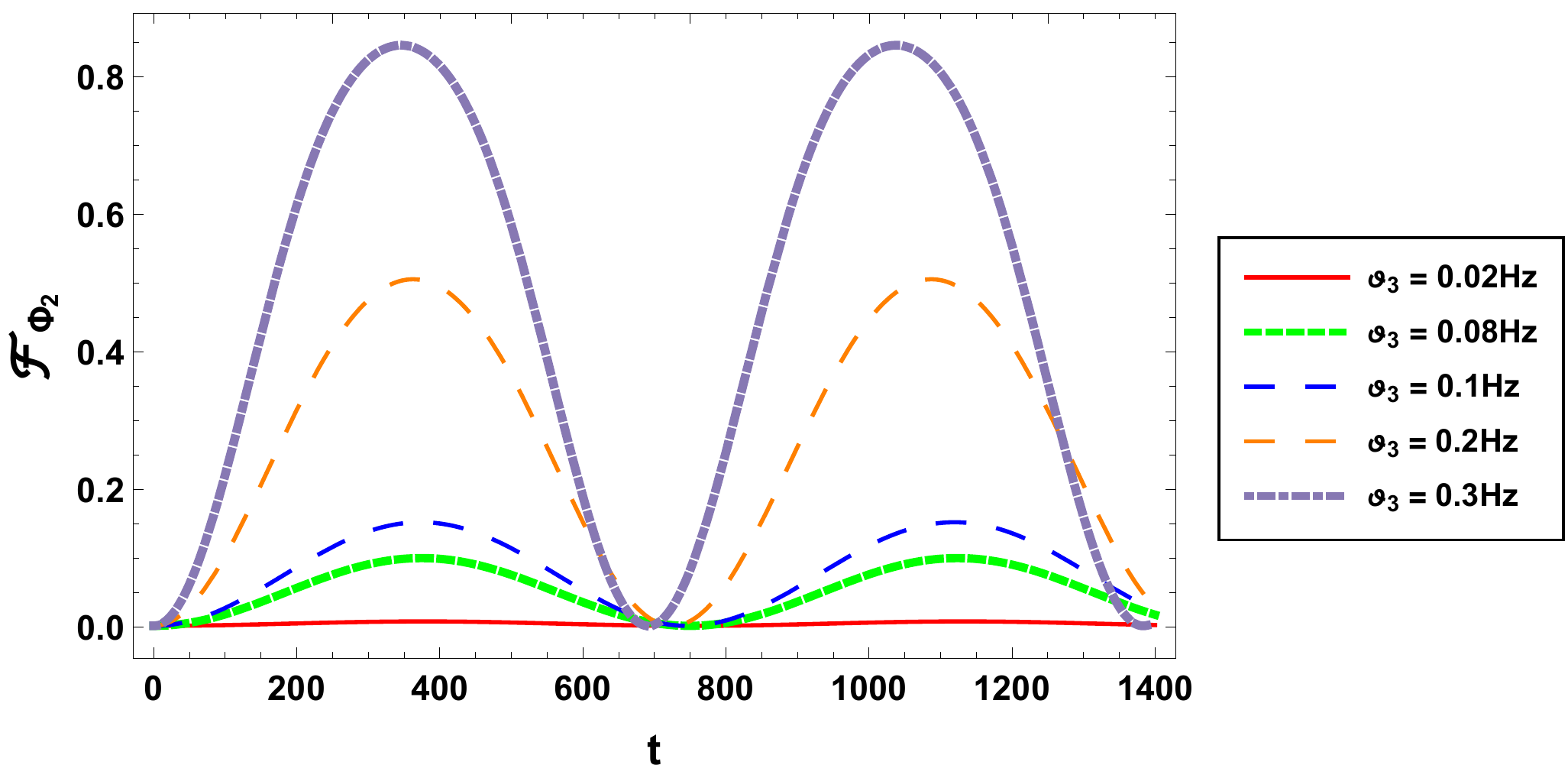} \vfill  $\left(b\right)$
				\end{minipage}\\
				\begin{minipage}[b]{.5\linewidth}
					\centering
					\includegraphics[scale=0.4]{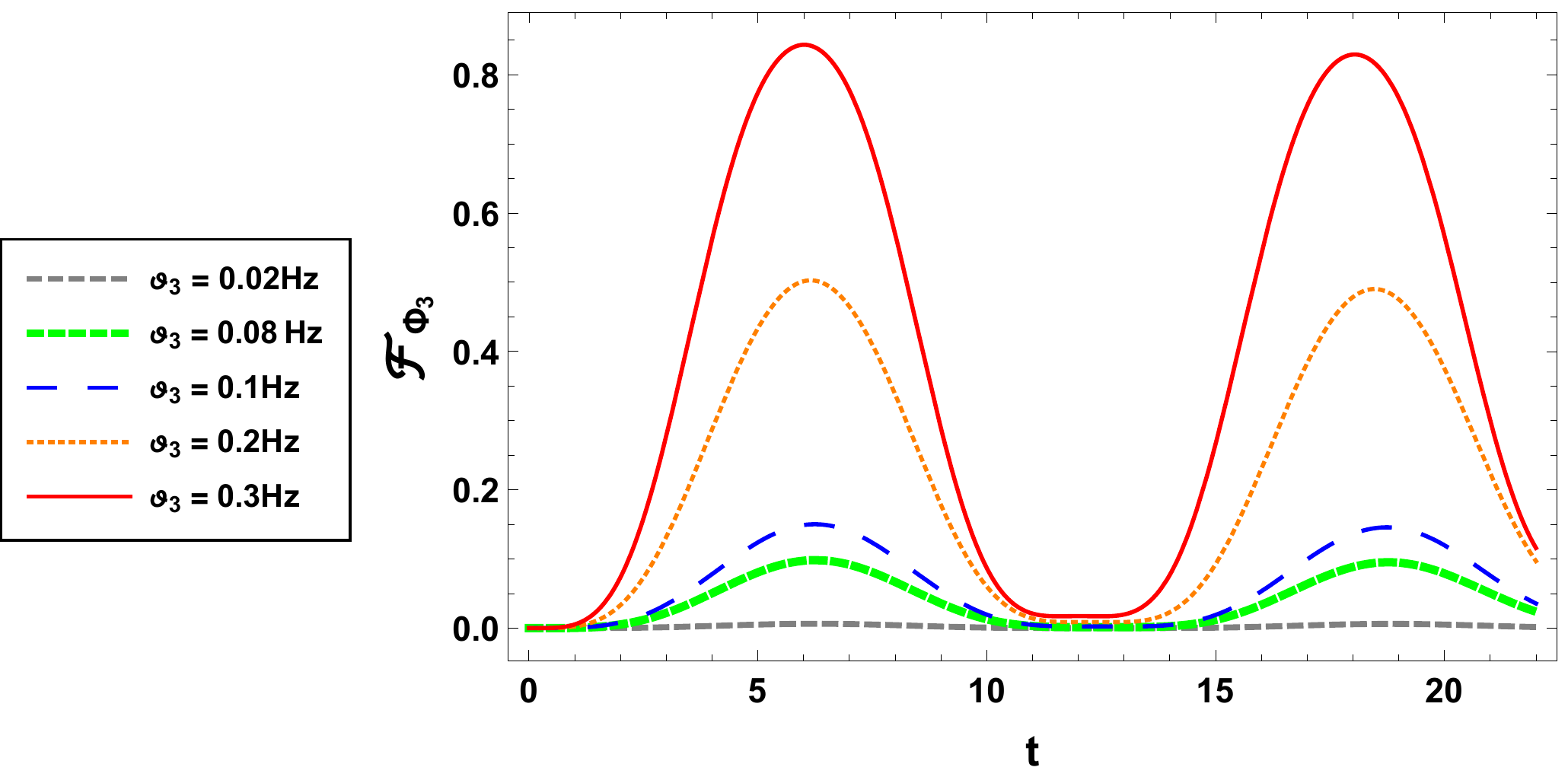} \vfill $\left(c\right)$
				\end{minipage}\hfill
				\begin{minipage}[b]{.5\linewidth}
					\centering
					\includegraphics[scale=0.4]{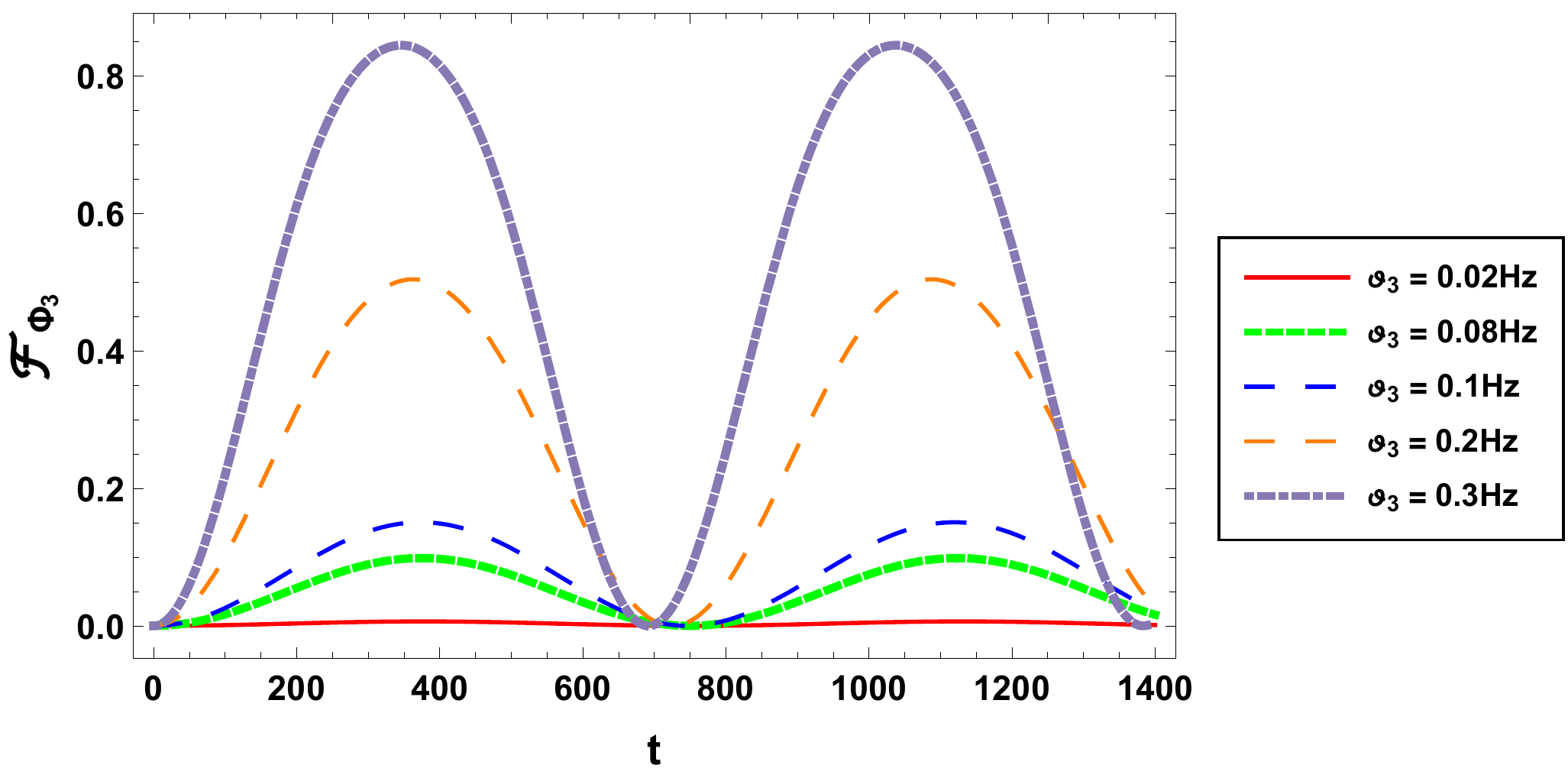} \vfill $\left(d\right)$
		\end{minipage}}}
		\caption{Dynamics of the single-parameter quantum Fisher information associated to the parameter $\phi_{2}$ and $\phi_{3}$ for different values of the dipole matrix element $\vartheta_{3}$ when $\vartheta_{2}=1$Hz and; $\Delta=0.04$Hz for panel ($a$) and panel ($c$), $\Delta=30$Hz for panel ($b$) and panel ($d$).} \label{Fig2}
	\end{figure}
\begin{figure}[hbtp]
	{{\begin{minipage}[b]{.5\linewidth}
				\centering
				\includegraphics[scale=0.4]{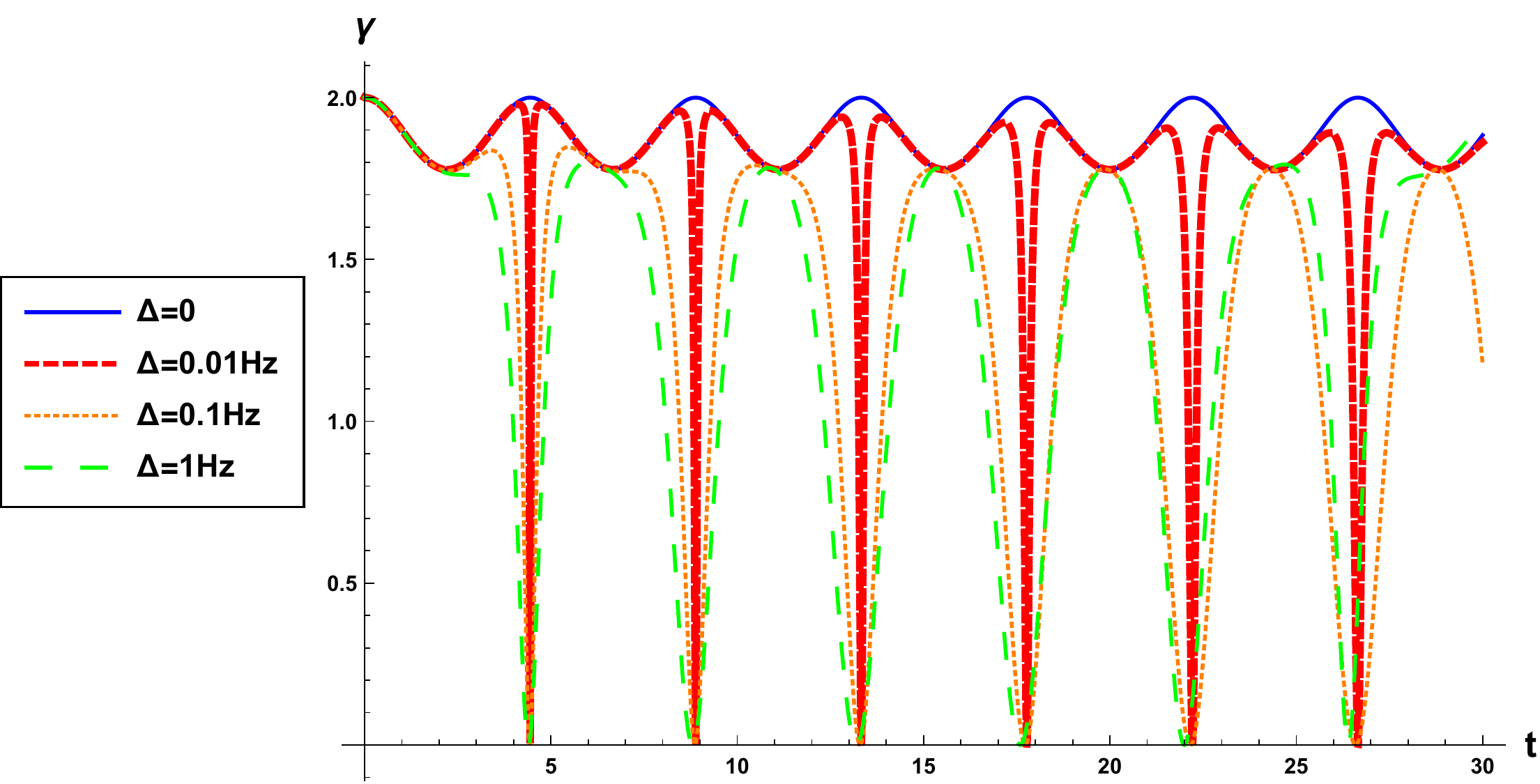} \vfill $\left(a\right)$
			\end{minipage}\hfill
			\begin{minipage}[b]{.5\linewidth}
				\centering
				\includegraphics[scale=0.45]{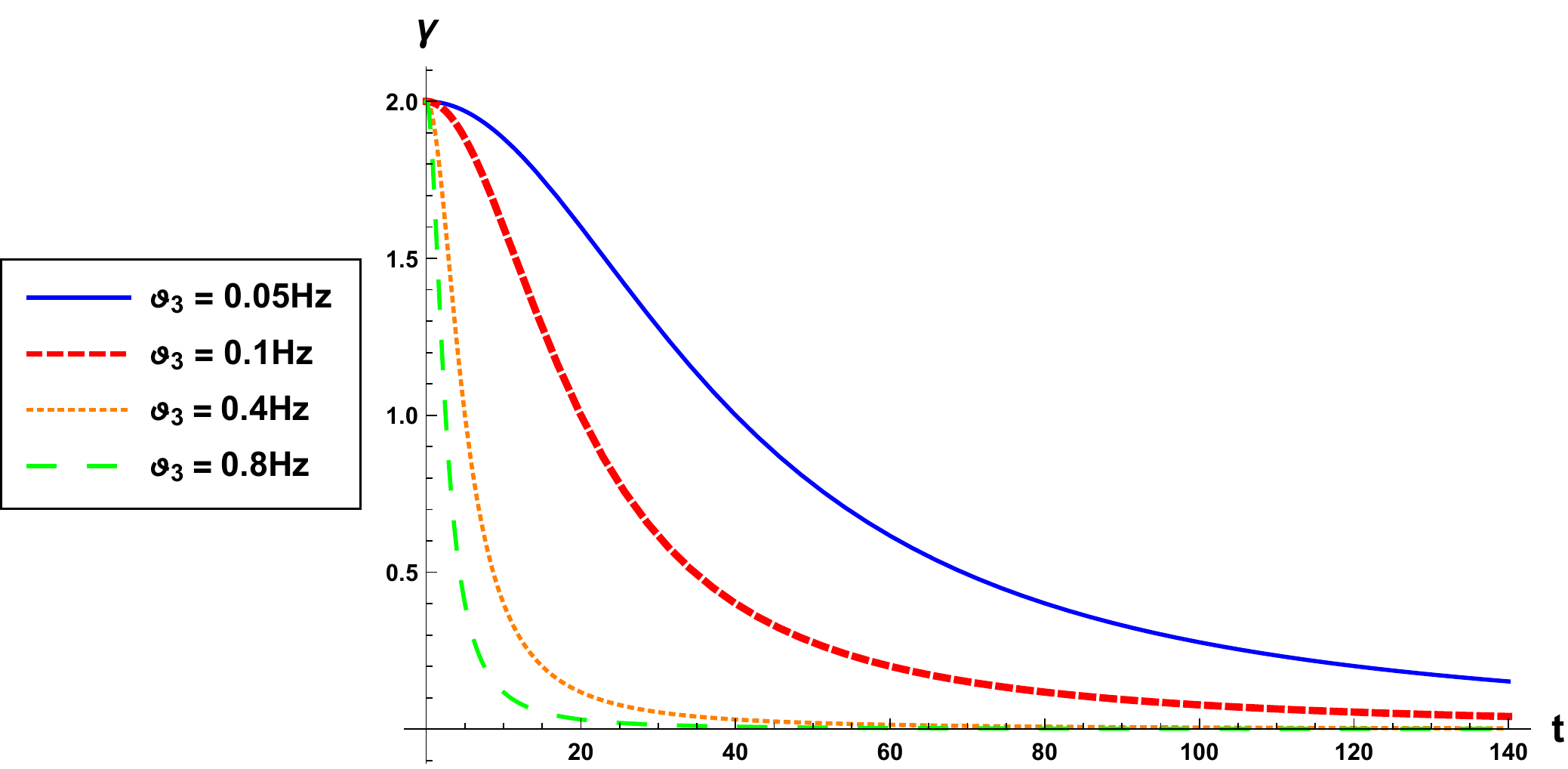} \vfill  $\left(b\right)$
			\end{minipage}}}
	\caption{Dynamics of the performance ratio $\gamma=\Delta_{\rm Ind}/\Delta_{\rm Sim}$ of the phase parameters to be estimated $\phi_{2,3}$; panel (a) for various values of the detuning parameter $\Delta$ with the dipole matrix elements are fixed as $\vartheta_{2}=\vartheta_{3}=1$Hz, panel (b) for various values of the parameter $\vartheta_{3}$ with $\vartheta_{2}=0.5$Hz and detuning parameter is set to $\Delta=30$Hz.} \label{Fig3}
\end{figure}
		\end{widetext}
The results reported in Fig.\ref{Fig2} illustrate the dynamics behavior of single-parameter quantum Fisher information related to the phase parameters $\phi_{2,3}$ for various values of the dipole matrix element $\vartheta_{3}$ when the three-level atom is initially prepared in its lower bare state. Remarkably, it is intriguing to notice that classical light fields have a high ability to reducing the error, captured by QFI, on the estimated phase parameters. Indeed, QFI has an oscillatory behavior with the same amplitudes and which increases with the increasing of the elements $\vartheta$. Returning to Eq.(\ref{eq4}) in which $\vartheta_{2,3}\propto E_{2,3}$, we conclude that the total variances of the estimated phase parameters are minimal, and therefore the estimation error is reduced, when the classical monochromatic fields increase. Moreover, the QFI vanishes for $t=0$ showing that the system evolves to a phase-sensitive state and no information on the phase parameters can be extracted from the three-level atom. Thereafter, the two classical monochromatic fields leads to the generation of QFI. This generated QFI survives during a certain time interval. Then, the amount of QFI decreases after which it increases again until it reaches its maximum. Moreover, increasing the dipole matrix element $\vartheta_{3}$ leads to enhancing the amount of QFI contained in the system. This implies that the best estimate of the phase parameters $\phi_{2,3}$ is achieved when the value of $\vartheta_{3}$ is large. Importantly, by comparing Fig.\ref{Fig2}($a$) with Fig.\ref{Fig2}($b$) for the estimated parameter $\phi_{2}$ and Fig.\ref{Fig2}($c$) with Fig.\ref{Fig2}($d$) for the estimated parameter $\phi_{3}$, we observe that the oscillatory amplitudes increase, thus the state is phase-insensitive, and a large amount of QFI could be preserved with the increase of the detuning parameter. Therefore, the large detuning parameters (where any three-level atom configuration includes two of these parameters and its third one has disappeared) helps to improve the precision of quantum estimation and it is important to use the three-level atom with high values of $\Delta$ to obtain the best estimation efficiency of the estimated parameters.\par

To evaluate the performance of metrological strategies, we display the performance ratio between the simultaneous and individual estimation schemes versus the time $t$ for various values of the detuning parameter in Fig.\ref{Fig3}($a$), and for various values of the parameter $\vartheta_{3}$ in Fig.\ref{Fig3}($b$). As shown in these Figures, the performance ratio takes its maximum value at $t=0$, which is clear that no advantage is attained by estimating the phase parameters individually. Also, the performance of simultaneous estimation is better and would provide more precise result than that of individual estimation for small values of both the parameters $\Delta$ and $\vartheta_{3}$. As depicted in Fig.\ref{Fig3}($a$), we remark that increasing the detuning parameter decreases the performance ratio and thus the superiority of the individual scheme over all parameter space. In Fig.\ref{Fig3}($b$), we visualize the influence of the dipole matrix element $\vartheta_{3}$ on the performance ratio $\gamma$ by taking $\vartheta_{2}=0.5$ and $\vartheta_{3}=30$. Here, the increase of the dipole matrix elements of the three-level atom leads to a rapid decrease in the ratio between the minimal total variances, indicates that the performance of the individual estimation is better than the simultaneous estimation.\par 

Our results show that the detuning and dipole matrix elements resulting from applying the two classical monochromatic fields can improve or reduce the precision of the parameters to be estimated. In other words, the optimal estimation efficiency and the performance of simultaneous and individual metrological strategies can be controlled by adjusting these two parameters. Accordingly, with the desirable choice, one can enhance or annihilate the estimation accuracy of the laser phases in the system.

\section{Dynamics of relevant quantum criteria and their role in quantum estimation}\label{Sec4}
In this section, we turn our attention on the evolution of the quantum coherence (QC) and the Hilbert-Schmidt speed (HSS) for the considered model (\ref{H}) as well as on their role in the phase estimation by comparing their dynamical behaviors with those of the QFI.
\subsection{\bf Quantum Coherence}
As a significant resource in quantum information science, QC resulting from quantum superposition is a key feature of quantum mechanics and it is crucial for a variety of quantum effects. In fact, coherent superposition of states is a key property of quantum systems and a necessary condition for quantum entanglement and other quantum resources \cite{Nielsen2000,Glauber1963,Streltsov2015}. Hence, QC directly affects the efficiency and reliability of quantum information processing. A rigorous framework for QC resource theory was recently introduced by Baumgratz and his colleagues \cite{Baumgratz2014} and some bona fide measures that satisfy all known criteria for a given quantum state have been put forward \cite{Girolami2014,Streltsov2017}. Several quantifiers are significant distance-based measures between the considered state $\rho$ and the nearest inconsistent state $\sigma$, i.e. $C_{D}\left(\rho\right)=\min_{\sigma\in I}D\left(\rho,\sigma\right)$ with $I$ is the set of all incoherent states, and the overall form of its distance is $D\left(\rho,\sigma\right)=\parallel\rho-\sigma\parallel$ where $\parallel.\parallel$ being some kind of norm matrix. In this paper, we mainly consider the $l_{2}$-norm of coherence as a measure of QC. It is defined as
\begin{equation}
C_{l_{2}}\left(\rho\right)=\min_{\sigma\in I}\parallel\rho-\sigma\parallel^{2}_{l_{2}}=\sum_{i,j}^{i\neq j}|\rho_{ij}|^{2}.
\end{equation}
Through calculation, in our theoretical model characterized by the atomic reduced density matrix (\ref{eq17}), we can achieve the analytical expression of $l_{2}$-norm of coherence as follows
\begin{equation}
C_{l_{2}}\left(\rho\right)=2\left(|\varepsilon\varkappa|^{2}+|\varepsilon\varpi|^{2}+|\varkappa\varpi|^{2}\right),
\end{equation}
with the variables $\varepsilon$, $\varkappa$ and $\varpi$ are given by
\begin{align}
&\varepsilon=2\vartheta_{2}\left(\omega_{i}W_{i}+W_{k}\omega_{k}\right),\notag\\&\varkappa=\vartheta_{2}^{2}\left(W_{i}+W_{k}\right)+\vartheta_{3}^{2}/(\vartheta_{2}^{2}+\vartheta_{3}^{2}),\notag\\&\varpi=\vartheta_{2}\vartheta_{3}\left(W_{i}+W_{k}-1/(\vartheta_{2}^{2}+\vartheta_{3}^{2})\right),
\end{align}
where
\begin{equation}
W_{i}=\frac{e^{-i\omega_{i}t}}{4\omega_{i}^{2}+\vartheta_{2}^{2}+\vartheta_{3}^{2}},\hspace{0.5cm}{\rm and}\hspace{0.5cm} W_{k}=\frac{e^{-i\omega_{k}t}}{4\omega_{k}^{2}+\vartheta_{2}^{2}+\vartheta_{3}^{2}}.
\end{equation}
\begin{figure}[hbtp]
	\centering
	\includegraphics[scale=0.4]{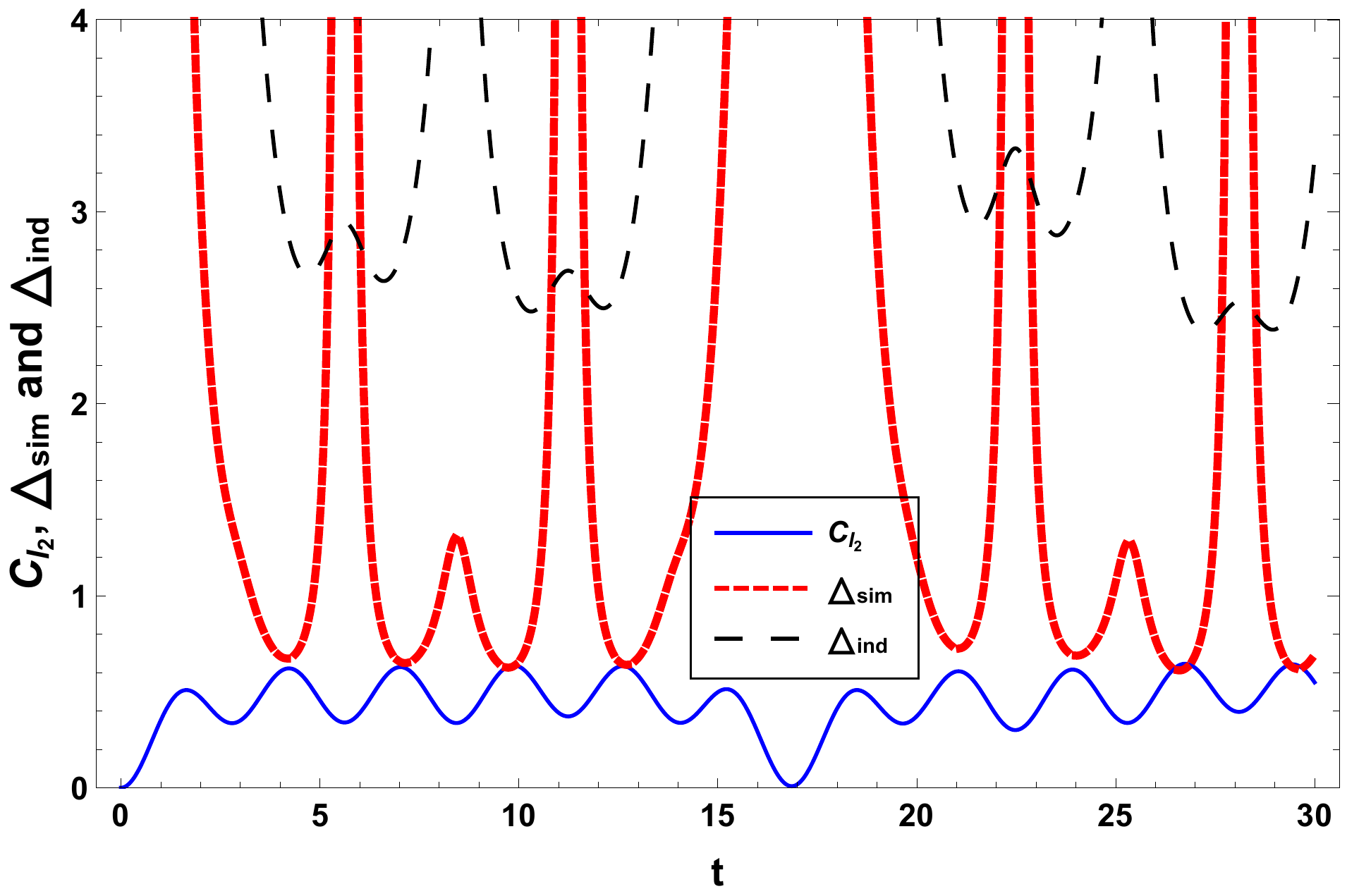}
	\caption{Dynamics comparison of quantum coherence (solid blue curve), the total variance in the individual (black dashed curve) and simultaneous (red dotted curve) schemes; here, $\vartheta_{2}=1$Hz, $\vartheta_{3}=0.3$Hz and $\Delta=0.04$Hz.}\label{Fig4}
\end{figure}

\subsection{\bf Hilbert-Schmidt Speed} 
We analyze here the HSS \cite{Gessner2018}, as a special type of quantum statistical speed, and see how it can be related to the QFI in order to improve the estimation of the quantum phase. Actually, HSS is recognized as an efficient method to identifying non-Markovianity in $d$-qubit open quantum systems \cite{Jahromi2020}. Consider first the class of distance measures between the two probability distributions $p=\{p_{x}\}_{x}$ and $q=\{q_{x}\}_{x}$ that are described as
\begin{equation}
|d_{\alpha}(p,q)|^{\alpha}=\frac{1}{2} \sum_{x}|p_{x}-q_{x}|^{\alpha},\label{eq46}
\end{equation}
where $\alpha\geq1$. To derive the statistical speed from any given statistical distance, we start by parameterizing the probability distribution $p\left(\theta\right)$ and then extend it to first order in $\theta$ at $\theta_{0}$, that is,
\begin{equation}
p_{x}\left(\theta_{0}+\theta\right)=p_{x}\left(\theta_{0}\right)+\frac{\partial p_{x}\left(\theta\right)}{\partial\theta}|_{\theta=\theta_{0}}\theta+{\cal O}\left(\theta^{2}\right).
\end{equation}
Employing this expansion, we obtain
\begin{equation}
d_{\alpha}\left(p\left(\theta_{0}+\theta\right),p\left(\theta_{0}\right)\right)=\left(\frac{1}{2}\sum_{x}|p'_{x}\left(\theta_{0}\right)|^{\alpha}\right)^{\frac{1}{\alpha}}\theta+{\cal O}\left(\theta^{2}\right),
\end{equation}
where $p'_{x}\left(\theta_{0}\right)=\partial_{\theta} p_{x}\left(\theta\right)|_{\theta=\theta_{0}}$. The classical statistical speed associated with the classical distance (\ref{eq46}) is given by
\begin{align}
s_{\alpha}\left[p\left(\theta_{0}\right)\right]&=\partial_{\theta}\left[d_{\alpha}\left(p\left(\theta_{0}+\theta\right),p\left(\theta\right)\right)\right]\notag\\&=\left(\frac{1}{2}\sum_{x}|p'_{x}(\theta_{0})|^{\alpha}\right)^{\frac{1}{\alpha}}.
\end{align}
When extending these classical notions to the quantum case and considering a given pair of quantum states $\rho$ and $\sigma$, then write $p_{x}={\rm Tr}\{E_{x}\rho\}$ and $q_{x}={\rm Tr}\{E_{x}\sigma\}$ which denote the measurement probabilities corresponding to the POVMs. The associated quantum distance can be obtained by maximizing the classical distance $d_{\alpha}(p,q)$ over all possible POVM choices. It is defined as \cite{Gessner2018}
\begin{equation}
D_{\alpha}(\rho,\sigma):=\max_{\{E_{x}\}}d_{\alpha}(p,q)=\left(\frac{1}{2}{\rm Tr}|\rho-\sigma|^{\alpha}\right)^{\frac{1}{\alpha}}.
\end{equation}
Therefore, we get the quantum statistical speed in the following form
\begin{equation}
S_{\alpha}\left[\rho_{\theta}\right]=\max_{\{E_{x}\}}s_{\alpha}\left[p\left(\theta\right)\right]=\left(\frac{1}{2}Tr|\partial_{\theta}\rho_\theta|^{\alpha}\right)^{\frac{1}{\alpha}},
\end{equation}
and for the situation where $\alpha=2$, the quantum statistical speed reduces to the Hilbert-Schmidt speed as
\begin{equation}
HSS\left[\rho_{\theta}\right]=HSS_{\theta}=S_{2}\left[\rho_{\theta}\right]=\sqrt{\frac{1}{2}{\rm Tr}|\partial_{\theta}\rho_\theta|^{2}},\label{eq52}
\end{equation}
which do not necessitate the diagonalization of $\partial_{\theta}\rho_\theta$. For a three-level atom interacting with two classical monochromatic fields, the analytical expression of the HSSs associated with the estimated phase parameters $\phi_{2,3}$ is obtained by applying equation (\ref{eq52}) to equation (\ref{eq17}). The result is
\begin{equation}
HSS_{\phi_{2}}=\sqrt{\varLambda\left(\Theta+\Sigma\right)}, \hspace{0.5cm} HSS_{\phi_{3}}=\sqrt{\Sigma\left(\varLambda+\Theta\right)},
\end{equation}
with the quantities $\varLambda$, $\Theta$ and $\Sigma$ are given by
\begin{align}
\varLambda= &\frac{\vartheta_{2}^{4}}{(4w_{i}^{2}+\vartheta_{2}^2+\vartheta_{3}^2)^{2}}+\frac{\vartheta_{2}^{4}}{(4w_{k}^{2}+\vartheta_{2}^2+\vartheta_{3}^2)^{2}} \notag \\
&+\frac{2\vartheta_{2}^{4}\cos((w_{i}-w_{k})t)}{(4w_{i}^{2}+\vartheta_{2}^2+\vartheta_{3}^2)(4w_{k}^{2}+\vartheta_{2}^2+\vartheta_{3}^2)}+\left(\frac{\vartheta_{3}^{2}}{\vartheta_{2}^{2}+\vartheta_{3}^{2}}\right)^{2}\notag \\
&+\frac{2\vartheta_{2}^{2}\vartheta_{3}^{2}}{\vartheta_{2}^{2}+\vartheta_{3}^{2}}\left(\frac{\cos(w_{i}t)}{4w_{i}^{2}+\vartheta_{2}^{2}+\vartheta_{3}^{2}}+\frac{\cos(w_{k}t)}{4w_{i}^{2}+\vartheta_{2}^{2}+\vartheta_{3}^{2}}\right),
\end{align}
\begin{align}
\Theta= &\frac{\omega_{i}^{2}}{(4\omega_{i}^{2}+\vartheta_{2}^{2}+\vartheta_{3}^{2})^{2}}+\frac{\omega_{k}^{2}}{(4\omega_{k}^{2}+\vartheta_{2}^{2}+\vartheta_{3}^{2})^{2}} \notag \\& +\frac{2\omega_{i}\omega_{k}\cos((\omega_{i}-\omega_{k})t)}{(4\omega_{i}^{2}+\vartheta_{2}^{2}+\vartheta_{3}^{2})(4\omega_{k}^{2}+\vartheta_{2}^{2}+\vartheta_{3}^{2})},
\end{align}
and
\begin{align}
\Sigma= &\frac{1}{(4\omega_{i}^{2}+\vartheta_{2}^{2}+\vartheta_{3}^{2})^{2}}+\frac{1}{(4\omega_{k}^{2}+\vartheta_{2}^{2}+\vartheta_{3}^{2})^{2}} \notag \\&+\frac{2\cos((\omega_{i}-\omega_{k})t)}{(4\omega_{i}^{2}+\vartheta_{2}^{2}+\vartheta_{3}^{2})(4\omega_{k}^{2}+\vartheta_{2}^{2}+\vartheta_{3}^{2})}+\frac{1}{(\vartheta_{2}^{2}+\vartheta_{3}^{2})^{2}}\notag \\& \quad -\frac{2}{\vartheta_{2}^{2}+\vartheta_{3}^{2}}\left(\frac{\cos(\omega_{i}t)}{4\omega_{i}^{2}+\vartheta_{2}^{2}+\vartheta_{3}^{2}}+\frac{\cos(\omega_{k}t)}{4\omega_{k}^{2}+\vartheta_{2}^{2}+\vartheta_{3}^{2}}\right).
\end{align}
To see the role of quantum coherence in the quantum estimation error boundary, the evolution of $l_{2}$-norm of coherence, total variance in the individual and simultaneous schemes are plotted in Fig.\ref{Fig4}. It is found from Fig.\ref{Fig4} that all these quantities exhibit a periodic behavior during the time evolution. Interestingly, the $l_{2}$-norm of coherence reaches the maximum possible value when the total variances in both metrological strategies are minimal. This implies that there is a relation between the quantum coherence and the total invariances of the estimated parameters where the decrease of quantum coherence corresponds to the growth of total invariances and the maximum total invariances corresponds to the minimum quantum coherence. Based on these results, a higher precision in all metrological strategies is achieved for the highest value of quantum coherence; this is in accordance with the results given by quantum entanglement \cite{SlaouiB2019,Huang2016,Slaoui2018} where the maximum entanglement of the probe states constitutes a key feature to obtain an optimal multi-parametric estimation.

\begin{widetext}
	
	\begin{figure}[hbtp]
		{{\begin{minipage}[b]{.25\linewidth}
					\centering
					\includegraphics[scale=0.28]{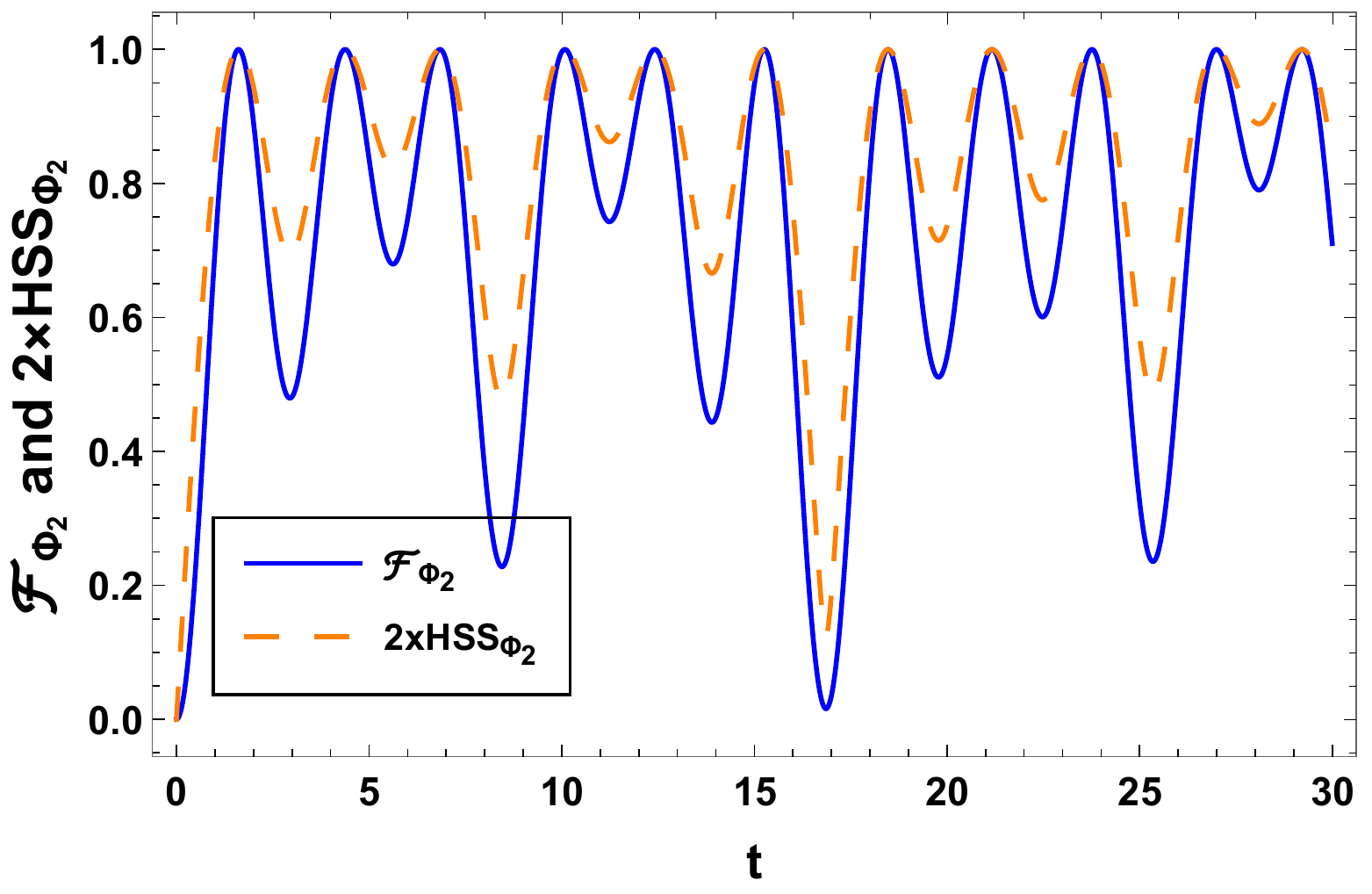} \vfill $\left(a\right)$
				\end{minipage}\hfill
				\begin{minipage}[b]{.25\linewidth}
					\centering
					\includegraphics[scale=0.28]{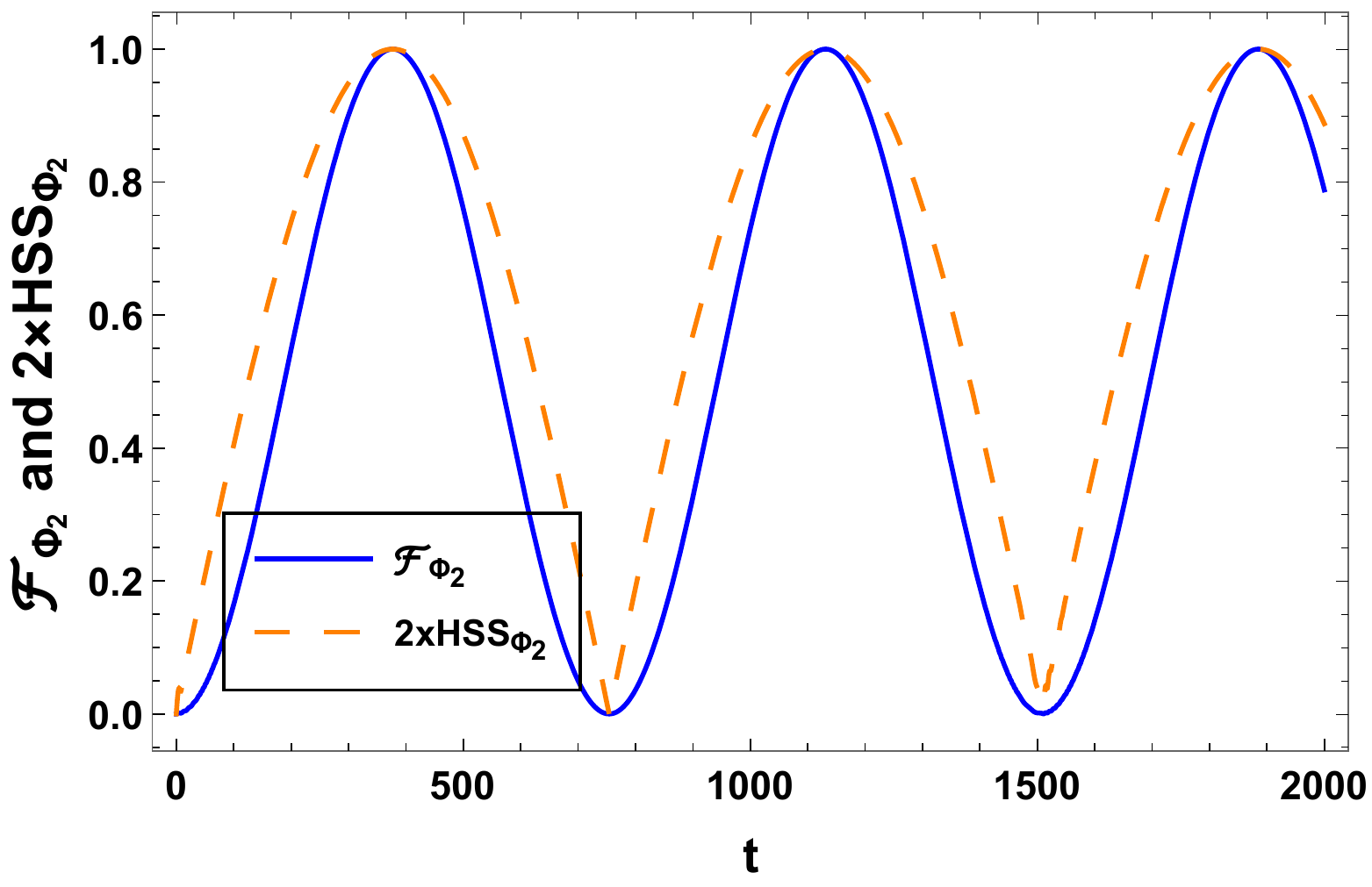} \vfill  $\left(b\right)$
				\end{minipage}\hfill
				\begin{minipage}[b]{.25\linewidth}
					\centering
					\includegraphics[scale=0.28]{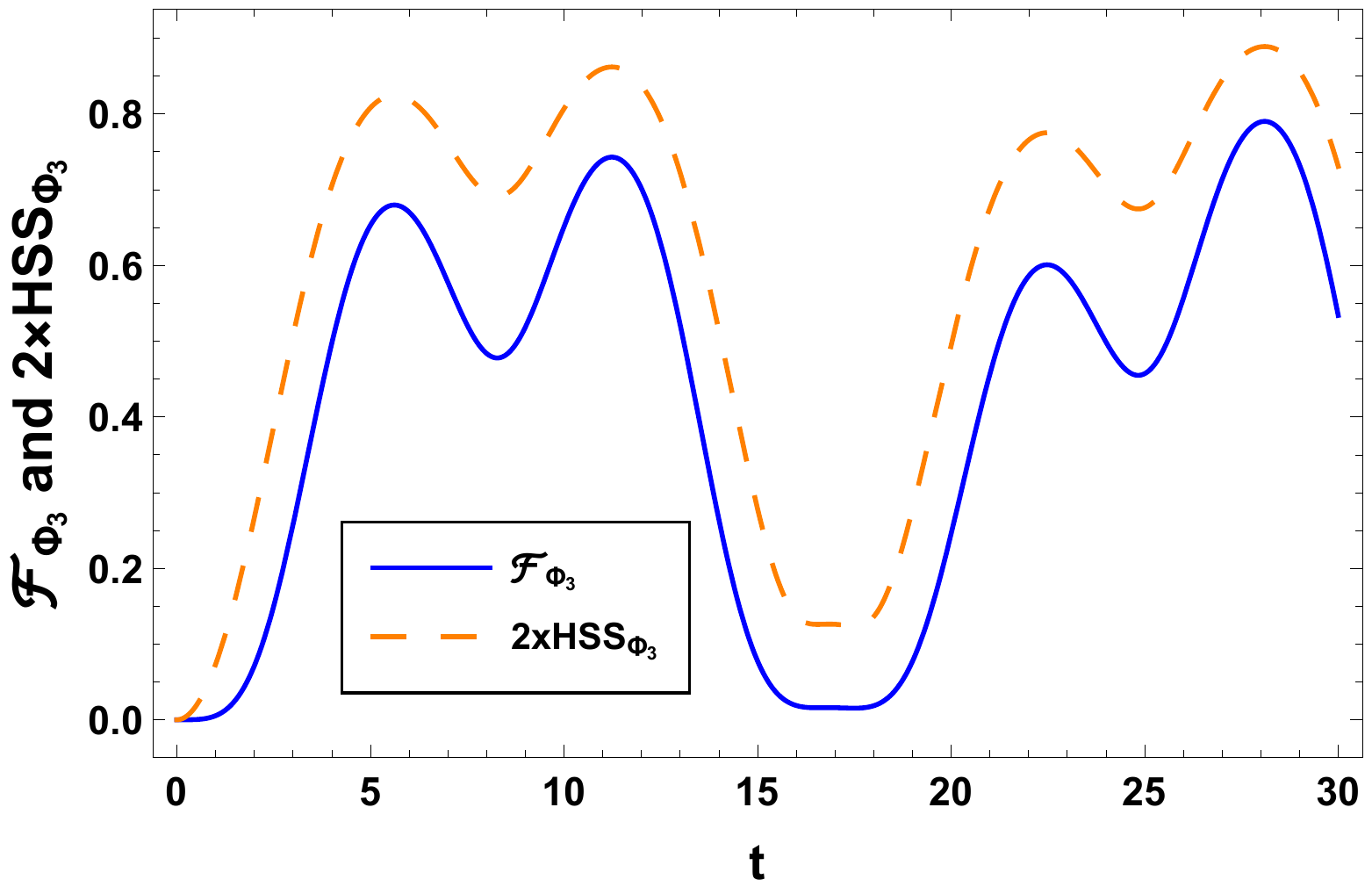} \vfill $\left(c\right)$
				\end{minipage}\hfill
				\begin{minipage}[b]{.25\linewidth}
					\centering
					\includegraphics[scale=0.28]{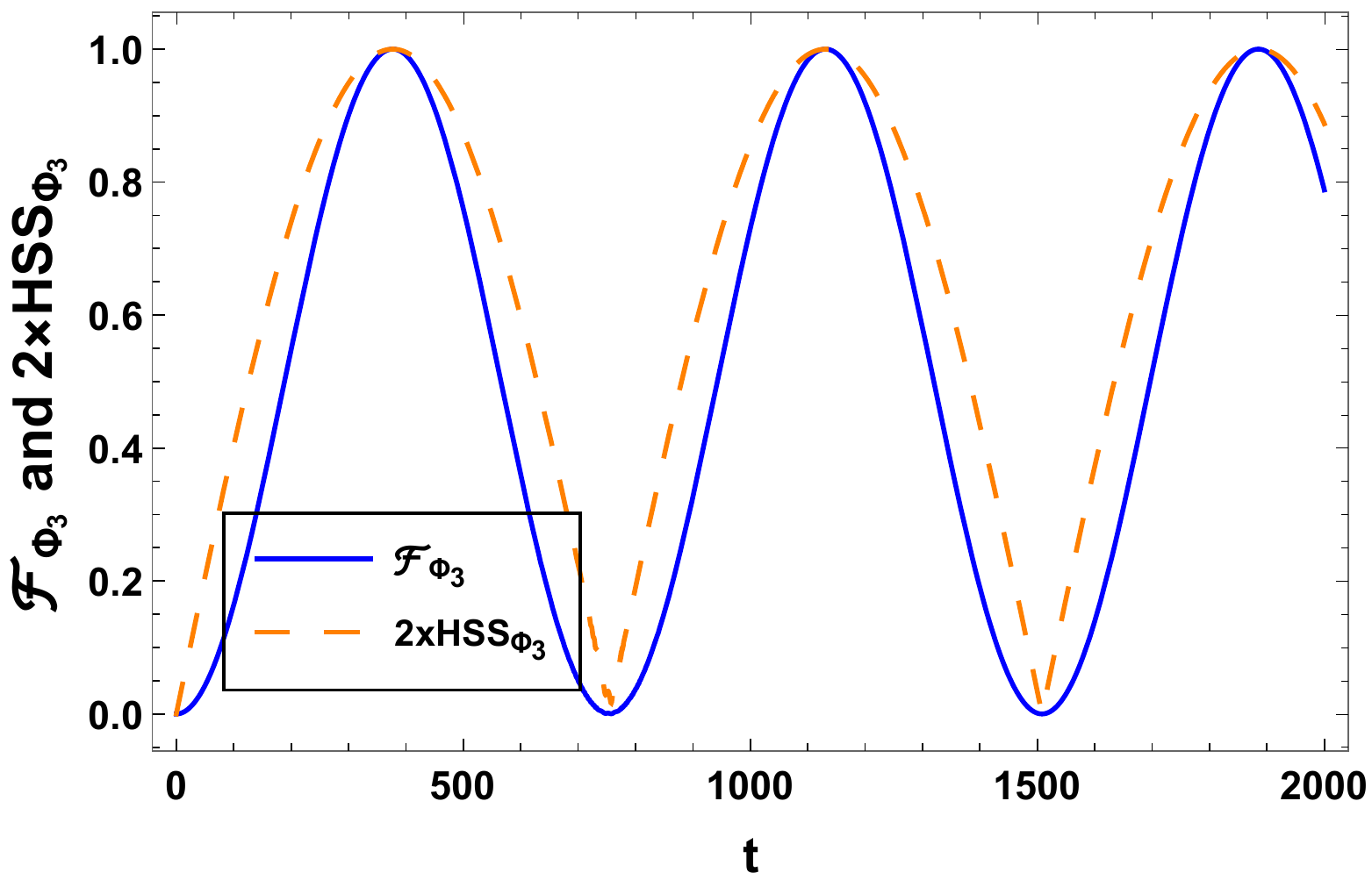} \vfill $\left(d\right)$
		\end{minipage}}}
		\caption{The comparison between the dynamics of the Hilbert-Schmidt speed (orange dashed curve) and the quantum Fisher information (solid blue curve) associated with the estimated phase parameters $\phi_{2,3}$; for panel ($a$) and panel ($c$) we fixed $\vartheta_{2}=1$Hz, $\vartheta_{3}=0.3$Hz and $\Delta=0.04$Hz; for panel ($b$) and panel ($d$) we take $\vartheta_{2}=\vartheta_{3}=0.5$Hz and $\Delta=30$Hz.}\label{Fig5}
	\end{figure}
\end{widetext}

In order to quantitatively compare HSS and QFI related to the laser phase parameters, we plot in Fig.\ref{Fig5} their evolutions versus the time $t$ for dipole matrix elements are set to $\vartheta_{2}=1$, $\vartheta_{3}=0.3$ and the detuning parameter is adjusted to $\Delta=0.04$ in Fig.\ref{Fig5}(($a$) for phase $\phi_{2}$ and ($c$) for phase $\phi_{3}$) and for $\vartheta_{2}=\vartheta_{3}=0.5$ and $\Delta=30$ in Fig.\ref{Fig5}(($b$) for phase $\phi_{2}$ and ($d$) for phase $\phi_{3}$). It is evident from Fig.\ref{Fig5} that both HSS and QFI dynamics exhibit oscillatory behaviors at the same time as well as their minimum and maximum locations aligning precisely. These results provide a qualitative indication that the HSS can be used to identify the precise times at which the best estimate of the estimated phase parameters occurs, corresponding to the maximum quantity of both HSS and QFI. Furthermore, our analysis has shown that the HSS can be employed as a practical and effective figure of merit to predict phase parameters, as it is a simple quantity to calculate and has the advantage of avoiding the diagonalization of the density matrix (\ref{eq17}) compared to QFI. These results are completely in agreement with the results reported in \cite{Rangani2021} which reflect that the HSS is a strong figure of merit to improve the quantum phase estimation in a multi-qubit quantum system. More broadly, it makes sense to investigate the links between these two concepts since they are quantum statistical speeds related to the Bures and Hilbert-Schmidt distances, respectively.

\section{Conclusion}
To conclude, the investigation of the three-level atomic system in various configurations interacting with a one- or two-mode field has received much attention in quantum optics. Using the rotating wave approximation, the atom-fields model has been extended to describe the idealized three-level atomic system. On another hand, quantum optical metrology addresses the estimation of an unknown phase by exploiting the quantum character of the considered input state. Its ultimate goal is to achieve a strong sensitivity of some probe states to small variations of external parameters as well as to find an ultimate measurement scheme to go beyond the standard quantum limits by which every quasi-classical estimation measurement is bounded, which opens great opportunities to increase the resolution of interferometric measurements.\par 

Here we summarize the important results obtained in this work. First, an exact solution of the considered theoretical model has been achieved which describes the interaction between a three-level atom and two classical monochromatic fields in the case of two-photon resonant transition, when the atom is initially prepared in the lower bare state taking into account the detuning parameters. Then, we investigated the multiparameter estimation strategy in the phase estimation protocol in the considered model by estimating two laser phases. We found their associated total variances, which gives us the optimal settings that effectively predict the values of the laser phase parameters, using the Cramér-Rao multiparameter quantum bound for simultaneous and individual estimation. Furthermore, the performance of the simultaneous parameter estimation was compared with the individual estimation by entering the ratio between the minimum amounts of total variances for each metrological strategies. The effect of detuning parameters and dipole matrix elements are explored, and our results suggest that the performance of simultaneous estimation was better and provided a more accurate result than individual estimation when both parameters had low values.\par

We also examine the roles of the Hilbert-Schmidt speed as well as the quantum coherence captured by the $l_{2}$-norm of coherence, contained in the three-level system, in improving the efficiency of quantum metrology protocols, particularly in multi-parameter unknown phase shift estimation protocols. We have shown that the Hilbert-Schmidt speed provides a powerful factor of merit for improving laser phases estimation, has an analytical computational advantage and testing its performance in large dimensional systems. Compared to the QFI, the HSS detects the regions where the estimation is optimal and plays the same role as the QFI in quantum metrology. Besides, quantum coherence assumes the same role as quantum entanglement in quantum estimation, whereby the maximum coherence coincides with the minimum variance of the estimated parameters, meaning that quantum coherence becomes an important resource for improving the sensitivity of quantum phase estimation. Building on the results obtained, a question that may naturally be arisen about the applicability of our analysis to other three-level atoms in various open-system interactions with the environment, including the atom-optomechanical system and many atom-cavity QED systems. In such models, the interaction between the probing system and its environment will be taken into account and the quantum advantage in phase detection in these real-world scenarios will be seriously affected. We hope to report on this issue in a forthcoming work.\par

Lastly, we hope that the current paper can help readers to investigate the three-level system in various tasks of quantum information theory, which may open new perspectives on quantum optical metrology.\\

{\bf Disclosures}: The authors declare no conflicts of interest or personal relationships that could have appeared to influence this work, and no Data associated in the manuscript.

\end{document}